\documentclass[useAMS,twocolumn,usenatbib]{mn2e}

\usepackage{graphicx,enumitem,color,hyperref}

\usepackage{amssymb,amsmath}

\newcommand{\lya}{Lyman-$\alpha$ }

\begin{document}

\title[The photoheating of the IGM]{The photoheating of the intergalactic medium in synthesis models of the UV background}
\author[E. Puchwein, J. S. Bolton, M. G. Haehnelt, P. Madau, G. D. Becker \& F. Haardt] {\parbox{\textwidth}{Ewald Puchwein$^{1}$, James S. Bolton$^{2}$, Martin
  G. Haehnelt$^{1}$,  Piero Madau$^{3,4}$,\\George D. Becker$^{1}$  and Francesco Haardt$^{5,6}$} \vspace{2mm}
  \\$^1$Institute of Astronomy and Kavli Institute for Cosmology, University of Cambridge, Madingley Road, Cambridge CB3 0HA, UK
  \\$^2$School of Physics and Astronomy, University of Nottingham, University Park, Nottingham, NG7 2RD, UK
  \\$^3$Department of Astronomy \& Astrophysics, University of California Santa Cruz, Santa Cruz, CA 95064, USA
  \\$^4$Center for Theoretical Astrophysics and Cosmology, Institute for Computational Science, University of Zurich, CH-9057 Zurich,\\ Switzerland
  \\$^5$DiSAT, Universit\`a dell'Insubria, Via Valleggio 11, I-22100 Como, Italy
  \\$^6$INFN, Sezione di Milano-Bicocca, Piazza delle Scienze 3, I-20123 Milano, Italy
}
\date{\today} 
\maketitle

\begin{abstract} 
We compare cosmological hydrodynamical simulations combined with the homogeneous
metagalactic UV background (UVB) of \citet{Haardt2012} (HM2012) to observations of the \lya forest
that are sensitive to the thermal and ionization state of the
intergalactic medium (IGM). The transition from optically thick to thin photoheating 
predicted by the simple one-zone, radiative transfer model implemented by 
HM2012 predicts a thermal history that is in remarkably good
agreement with the observed rise of the IGM temperature at $z\sim 3$ 
if we account for the expected evolution of the volume filling factor of He\,\textsc{III}.
Our simulations indicate that there may be, however, some tension between
the observed peak in the temperature evolution and the rather slow
evolution of the He\,\textsc{II} opacities suggested by recent Hubble Space Telescope/COS
measurements. The HM2012 UVB also underpredicts the metagalactic hydrogen photoionization rate 
required by our simulations to match the observed opacity of the forest at $z>4$ and $z<2$.
\end{abstract}


\section{Introduction}
\label{sec:introduction}

The thermal state of the intergalactic medium (IGM) at the moderate
overdensities probed by the \lya forest is generally believed to be
set by the balance of photoheating of hydrogen and helium by the metagalactic UVB
and adiabatic cooling/heating.  These competing effects result in a characteristic
temperature density relation of the low-density IGM
\citep{Hui1997,Valageas2002}. Shock heating as well as collisional
cooling processes also contribute, but these occur mainly at
overdensities larger than those probed by the \lya forest 
(defined here as absorption lines with column densities $N_{\rm HI}
\la 10^{14.5}\rm\,cm^{-2}$). The main uncertainties in this picture
arise from radiative transfer effects during the epoch of reionization, 
which are difficult to model accurately (\citealt{AbelHaehnelt1999}).

In the redshift range best accessible by observations of the
forest, $2<z<4$, photoheating during the reionization of
He\,\textsc{II} is expected to lead to an increase in the IGM 
temperature above that otherwise expected following the
completion of H\,\textsc{i} reionization at $z\ga 6$
(\citealt{Theuns2002,HuiHaiman2003}). There is a wide range of
evidence from \lya forest data to support heating from He\,\textsc{II} reionization at $z<6$
\citep{Schaye2000,Ricotti2000,BryanMachacek2000,Zaldarriaga2001,McDonald2001,Lidz2010,Becker2011,Garzilli2012,Bolton2012},
but see \citet{Puchwein2012} for possible additional contributions to
the observed heating.

Quantitative modelling of the \lya forest and the IGM at
$2<z<4$ in cosmological hydrodynamical simulations without radiative transfer
often relies on
boosting the photoheating rates predicted by homogeneous models of the UVB 
by factors of order two to account for non-equilibrium
ionization (e.g. \citealt{HaehneltSteinmetz1998,Theuns1998})
and radiative transfer effects during He\,\textsc{II} reionization. Such modified heating rates
have enabled these simulations -- which often also assume
photoionization equilibrium -- to match a particular observational
constraint by design (\citealt{Wiersma2009}).  Alternatively, a
range of rescaled heating rates may be used to marginalise results
over a plausible range of IGM temperatures (\citealt{Jena2005,Viel2013}).

In the last decade, the accuracy of measurements of the ionization
\citep[e.g.][]{Becker2013,Becker2014,Syphers2014,Worseck2014} and
thermal \citep[e.g.][]{Becker2011,Rudie2012,Bolton2014,Boera2014}
state of the IGM, as well as of cosmological parameter constraints
\citep{Planck2013} and synthesis modelling of the UVB
\citep[][hereafter HM2012]{Faucher-Giguere2009,Haardt2012}, have all
substantially improved. It appears then timely to investigate the extent 
to which the observations can be explained with accurate numerical 
modelling of the thermal and ionization state of the IGM. 
Cosmological hydrodynamical simulations with full radiative transfer
are still very challenging \citep{Paschos2007,McQuinn2009,MeiksinTittley2012,Compostella2013,Compostella2014},
and are not yet efficient enough to allow the exploration at the needed resolution 
of a large parameter space in boxes of size comparable to the mean free path 
of ionizing radiation. Here, we follow a hybrid approach, 
where we combine smooth particle hydrodynamics (SPH) simulations performed with a non-equilibrium ionization 
version of the \textsc{p-gadget3} code with state-of-the-art one-zone radiative transfer
calculations of a homogeneous, evolving UVB.

The paper is structured as follows. We describe the numerical methods 
in Sec.~\ref{sec:methods}, including a more detailed
  overview of the transition from optically thick to optically thin
  heating in the HM2012 UVB model. In
Sec.~\ref{sec:results} we present the main results regarding the
thermal state of the IGM in our simulations, and compare these
  predictions with the latest data from \lya
  forest observations. A discussion of our results in the
  context of previous radiative transfer simulations of the IGM
  thermal history at $2<z<4$ is presented in
  Sec.~\ref{sec:discussion}. Finally, we summarize our findings and
  conclude in Sec.~\ref{sec:conclusions}. In the
Appendices~\ref{sec:non-eq_effects} and \ref{sec:curv_vs_powspec}, we
elaborate on non-equilibrium ionization effects and the relation
between spectral curvature (\citealt{Becker2011}) and flux power
spectrum.

\section{Methodology}
\label{sec:methods}

\subsection{Hydrodynamical simulations}
\label{sec:sims}

Throughout this work, we make use of a set of cosmological
hydrodynamical simulations that were performed with the TreePM-SPH
simulation code \mbox{\sc p-gadget3}, an updated and significantly
extended version of {\sc gadget-2} \citep{Springel2005}. The
simulations adopt the best-fit \textit{Planck+lensing+WP+highL}
cosmology \citep{Planck2013} with $\Omega_{\rm M}=0.305$,
$\Omega_{\Lambda} = 0.695$, $\Omega_{\rm B}=0.0481$, $h=0.679$,
$\sigma_{8}=0.827$ and $n_s=0.962$. Runs with different box sizes and
resolutions were performed in order to assess the
numerical convergence of our results. An overview of all simulations
used in this work is provided in Table~\ref{tab:sims}.

\begin{table*}
\begin{tabular}{lcccccl}
\hline\hline
Simulation & Box size & $n_{\rm part}$ & $m_{\rm gas}$ & $m_{\rm dm}$ & $\epsilon$ & Photoheating models\\
name & $[\,h^{-1} \mathrm{Mpc}]$ & & $[\,h^{-1} M_\odot]$ & $[\,h^{-1} M_\odot]$ & $[\,h^{-1} \mathrm{kpc}]$\\
\hline\hline
L20N128  & 20 & $2\times128^3$ & $5.1\times10^7$ & $2.7\times10^8$ & 6.3 & modified HM1996 eq., HM2012 eq., HM2012 non-eq.,\\
&&&&&& HM2012 non-eq. no He \textsc{iii}, modified HM2012 non-eq.\\
\hline
L20N512  & 20 & $2\times512^3$ & $8.0\times10^5$ & $4.2\times10^6$ & 1.6 & HM2012 eq., HM2012 non-eq.\\
\hline
L10N512  & 10 & $2\times512^3$ & $9.9\times10^4$ & $5.3\times10^5$ & 0.78 & HM2012 non-eq., modified HM2012 non-eq.\\
\hline\hline
\end{tabular}
\caption{Summary of the parameters of the simulations used in this
  work. The columns list the simulation name,
  the comoving box size, the number of particles $n_{\rm
    part}$ in the initial conditions (half of which are gas and half dark matter 
    particles), the gas particle mass $m_{\rm
    gas}$, the dark matter particle mass $m_{\rm dm}$, the comoving
  gravitational softening $\epsilon$ (Plummer equivalent) and the different photoheating 
  models with which the simulations were
  performed (see main text for a detailed description).}
\label{tab:sims}
\end{table*}

All our simulations use a simplified model for star formation. All gas particles 
that exceed a density of 1000 times the mean baryon density and have a temperature 
below $10^5\rm\,K$ are converted to stars.  While this model results in unrealistic galaxy populations, 
it has been shown to yield the same
properties of the IGM at the relatively low densities probed by the
Lyman-$\alpha$ forest when compared to more sophisticated
star formation and feedback models \citep{Viel2004,Dave2010}. The simple scheme adopted here
is numerically much more efficient. 

\subsection{Equilibrium and non-equilibrium ionization}
\label{sec:eq}

As discussed in the Introduction, many numerical studies of
the Lyman-$\alpha$ forest include photoheating by assuming an IGM that 
is in photoionization equilibrium with an external homogeneous UVB. While 
photoionization equilibrium is a very good approximation after 
reionization, non-equilibrium effects during the reionization of hydrogen and helium 
can be significant.

In our analysis here, we compare simulations using the simplifying assumption of ionization equilibrium
(referred to as \textit{equilibrium} simulations) as well as more realistic simulations in which we drop this assumption 
(referred to as \textit{non-equilibrium} simulations). 
In both cases, the heating and cooling rate equations are solved time-dependently.
In equilibrium simulations, the ionization fractions of hydrogen
and helium are found using the method described in
\citet{Katz1996}. In non-equilibrium simulations, we integrate the ionization and recombination
rate equations (see e.g. appendix B3 in
\citealt{Bolton2007}). We follow \citet{Oppenheimer2013} in using the \textsc{CVODE}
library\footnote{http://computation.llnl.gov/casc/sundials/main.html}
\citep{Cohen1996,Hindmarsh2005} for this purpose. This is well suited
for efficiently integrating stiff ordinary differential equations
with variable-order, variable-step Backward Differentiation Formula
methods. In the non-equilibrium simulations, \textsc{CVODE} is used to evolve the
ionization states and the corresponding change in thermal energy of
the SPH particles to the next synchronization point, i.e. the next
time when particles require a force computation. Thus, for each
gravity/hydrodynamic simulation timestep and each SPH particle,
effectively a sub-cycling with a variable number of sub-steps is
performed for integrating the rate equations. Further details on
integrating the rate equations are given in
Appendix~\ref{sec:integ_rate_eq}. 

In both the equilibrium and non-equilibrium simulations, we take the case A recombination rates
from \citet{Verner1996}, the dielectric He \textsc{i} recombination
rate from \citet{Aldrovandi1973}, the collisional ionization rates
from \citet{Voronov1997}, the collisional excitation cooling rates
from \citet{Cen1992}, and the free-free bremsstrahlung cooling rate from
\citet{Theuns1998}. Throughout this work, we use the photoionization and photoheating rates from
HM2012 in our simulations (see their Table~3). There are only two exceptions. We perform one simulation (shown in Fig.~\ref{fig:T0}) with a modified \citet{Haardt1996} (hereafter HM1996) background
\citep{Dave1999}, i.e. {\sc p-gadget-3}'s default UVB file
\citep[see][]{Springel2003} to facilitate comparison to the literature. We perform another run with a modified version of the HM2012 UVB as will be discussed in Sec.~\ref{sec:curv_temp}.

\subsection{Photoionization and photoheating from the UVB}
\label{sec:hmuvb}

The updated, homogeneous, UVB of HM2012 is based on an empirically motivated model for the redshift evolution of the
spatially-averaged UV emissivity (of galaxies and quasars) and intergalactic opacity as a function of 
frequency. The background flux is obtained by solving a global radiative transfer equation in an expanding Universe,
\begin{equation}
  \left( \frac{\partial}{\partial t} - \nu H \frac{\partial}{\partial \nu} \right) J_\nu + 3 H J_\nu = -c \kappa_\nu J_\nu + \frac{c}{4 \pi} \epsilon_\nu, 
\label{eq:Jnu}
\end{equation}
where $J_\nu$, $H$, $c$, $\epsilon_\nu$, $\kappa_\nu$, $t$, and $\nu$ are 
the space- and angle-averaged monochromatic intensity, Hubble parameter, speed of light, proper volume emissivity, intergalactic absorption coefficient, 
cosmic time, and frequency, respectively. As it is the UVB that is responsible for the photoheating of the IGM on large scales, 
such a model should provide a realistic heating rate once the mean free path in the IGM is larger than the mean separation between the ionizing sources. 
The hydrogen and helium photoionization rates are given by 
\begin{equation}
  \Gamma_i = \int_{\nu_i}^\infty\textrm{d} \nu \, \frac{4 \pi J_\nu}{h \nu} \, \sigma_i(\nu),
\end{equation}
where $h$ is Planck's constant, the subscript $i$ denotes the relevant ion species, $h \nu_i$ is the ionization energy,
and $\sigma_i(\nu)$ is the photoionization cross section.
The ensuing spatially-uniform photoheating rate is given by 
\begin{equation}
 \mathcal{H}_i = \int_{\nu_i}^\infty \textrm{d} \nu \, \frac{4 \pi J_\nu}{h \nu} \, h (\nu - \nu_i) \, \sigma_i(\nu).
\label{eq:Hi}
\end{equation}
Note that the formula above provides the {\it correct} heating rate of intergalactic gas once the background intensity
$J_\nu$ is {\it properly filtered} while propagating through the IGM. The treatment of this spectral filtering in HM2012 is based on empirical constraints on the abundance of absorbers as a function of their H\,\textsc{i} column density $N_{\rm HI}$. Local radiative transfer models of the absorbers are used to relate this H\,\textsc{i} column density to the He\,\textsc{i} and He\,\textsc{ii} column densities of the absorber. Effectively, mean redshift and UV background-dependent one-to-one conversions between $N_{\rm HI}$ and the He\,\textsc{i} and He\,\textsc{ii} column densities $N_{\rm HeI}$ and $N_{\rm HeII}$ are used. This approach is certainly well motivated after the reionization of the considered species. Remaining uncertainties in the spectral filtering during He\,\textsc{ii} reionization are discussed in more detail in Appendix~\ref{sec:heii_absorbers}, as well as at in the last paragraph of this section. Further uncertainties during hydrogen reionization arise due to the extrapolation of the empirical absorber column density distribution to high redshifts $z \gtrsim 6.5$ where we lack observational constraints. Due to a ``loss-of-memory-effect'' caused by the subsequent cosmic expansion and photoheating \citep[see e.g.][]{HuiHaiman2003} this will not, however, affect IGM temperatures at the redshifts we are mostly concerned with in this work strongly.

\begin{figure}
\centerline{\includegraphics[width=\linewidth]{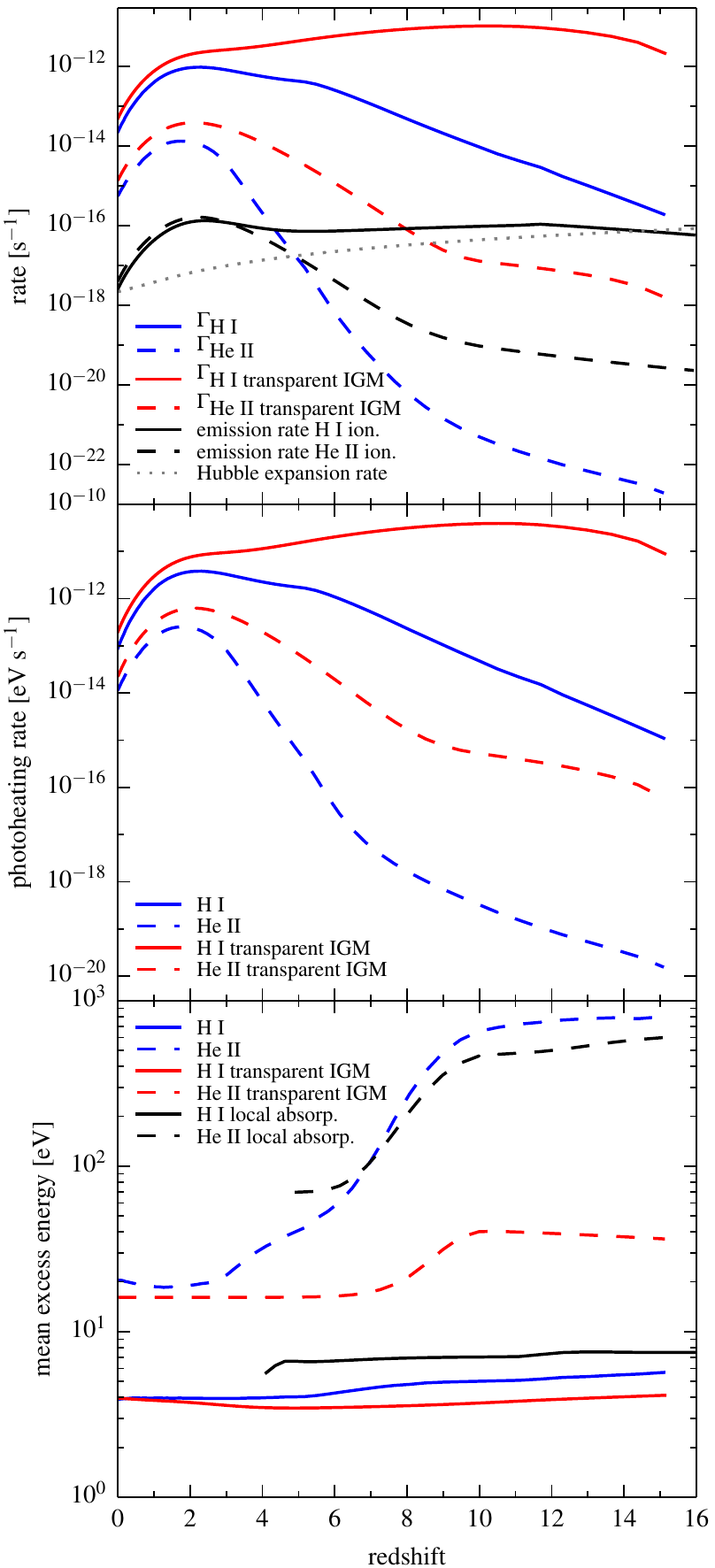}}
\caption{\textit{Top panel}: Ionizing photon emission and photoionization rates for 
  hydrogen and singly-ionized helium in the HM2012 model. For comparison, the 
  photoionization rates are also shown in the transparent IGM limit. 
  The emission rates of H\,\textsc{i}
  and He\,\textsc{ii} ionizing radiation are plotted per hydrogen and helium
  atom (including all ionization states), respectively. The Hubble expansion rate $H(z)$ is also indicated.
  \textit{Middle panel}: Photoheating rates for
  H\,\textsc{i} and He\,\textsc{ii} as a function of redshift in the HM2012 model as well as in the 
  transparent IGM limit. \textit{Lower panel}: Mean excess energy per
  H\,\textsc{i} and He\,\textsc{ii} ionization in the HM2012 model, as well as in
  the transparent IGM and immediate local absorption limits.
  }
\label{fig:uvb}
\end{figure}

Figure~\ref{fig:uvb} shows some of the relevant quantities predicted by the HM2012 model.\footnote{Comoving emissivities, 
background intensities, photoheating and photoionization rates are made available by Francesco Haardt and Piero Madau at this URL:
\url{http://www.ucolick.org/~pmadau/CUBA/DOWNLOADS.html}} 
The \textit{top panel} shows the assumed evolution of the photon emission rates above 1 and 4 Ry from galaxies and quasars, 
as well as the predicted hydrogen and singly ionized helium photoionization rates and the Hubble expansion rate, $H(z)$, corresponding 
to the adopted cosmology. Reionization occurs approximately when the photoionization rate exceeds the expansion and the radiative recombination
rates. The much larger increase with cosmic time of the photoionization rate compared to the emission rate is explained by an increase 
in the photon mean free path. In the \textit{middle panel} we plot the H\,\textsc{i} and He\,\textsc{ii} photoheating rates from Eq.~(\ref{eq:Hi}).
Note that we have not shown the He\,\textsc{i} rates. The reionization of neutral helium is completed at approximately the same time as
H\,\textsc{i} reionization, but it has a comparatively small effect on the thermal state of the IGM. 
Finally, the \textit{bottom panel} shows the excess energy per ionization of hydrogen and He\,\textsc{ii} as a function of redshift.

For illustrative purposes, we compare the ionization rate, photoheating rates, and excess energy obtained in the case of a HM2012 filtered 
UVB spectrum with the corresponding quantities derived under the assumption of negligible intergalactic opacity (i.e. $\kappa_\nu=0$ in Eq.~(\ref{eq:Jnu})).
Moreover, we display the excess energy in the opposite limit of immediate local absorption of ionizing photons (see Appendix~\ref{sec:loc_absorp} for details).
We shall refer to these limiting cases in the 
following as the {\it transparent IGM} and {\it local absorption} approximation, respectively. At low redshift, when the universe is transparent to ionizing radiation, the 
HM2012 heating rates are, as expected, close to the transparent IGM limit. By contrast, at high redshift the He\,\textsc{ii} heating rate is in good agreement with the local absorption approximation as the mean free path of $\gtrsim 4$ Ry photons is small compared to the Hubble radius. For H\,\textsc{i}, the mean excess energy also increases towards the local absorption expectation with increasing redshift, but does not fully reach it. The reason for falling short of the local absorption approximation even at 
early times is due to the softening of the background spectrum above the hydrogen ionization threshold associated 
with recombination radiation from the IGM (see e.g. Fig.~5c in HM1996).

We remark here that the one-zone radiative transfer calculations used by HM2012 
to generate the UVB follow the propagation of an external radiation field through slabs of hydrogen and helium gas
over a wide range of column densities, including the low columns associated with the forest. 
It is not clear whether such a spatially homogeneous UVB will correctly predict 
the characteristic photoheating rates during the epoch of reionization, when the low-density IGM makes the transition 
from neutral to highly ionized and the mean free path of ionizing radiation is typically much shorter
than the mean source separation. During this era different regions of the Universe will be subject 
to a UV flux whose amplitude and spectral shape depend on the distance to the nearest sources. 
There may also be systematic differences in the spectral shape of the radiation heating the IGM that
depend on (over-)density (\citealt{AbelHaehnelt1999,Bolton2004,Tittley2007}).
In the observationally best accessible range $2<z<4$, the transition from
He\,\textsc{ii} to He\,\textsc{iii} will be particularly problematic
in this regard \citep{Paschos2007,McQuinn2009,MeiksinTittley2012,Compostella2013,Compostella2014}. Nevertheless
as the hydrogen intergalactic opacity to ionizing radiation 
is well constrained by \lya forest and Lyman Limit system data, and the
modelling of helium absorption has been greatly improved by recent observations 
(e.g. \citealt{Syphers2014,Worseck2014}), it is worthwhile 
to investigate the thermal history of the IGM predicted by hydrodynamical simulations that
assume a spatially-averaged homogeneous UVB.

\subsection{Synthetic Lyman-$\alpha$ forest spectra}
\label{sec:spectra}

We compute synthetic Lyman-$\alpha$ forest spectra in post-processing. This allows us to 
directly compare the effective optical depth for absorption as well as other statistics of
the simulated spectra to observations.  We select 5000 randomly placed lines of
sight through each output of the simulation box, along directions
parallel to one of the coordinate axes (randomly selected among $x$,
$y$, and $z$). Each line of sight is represented by 2048 pixels. We
then compute the neutral hydrogen density, temperature, and velocity of
the IGM along these lines of sight by adding up the density
contributions and averaging the temperatures and velocities of all SPH
particles whose smoothing lengths are intersected. Our calculation of
the spectra accounts for Doppler shifts due to bulk flows of the gas
as well as for thermal broadening of the Lyman-$\alpha$ line
\citep[see e.g.][]{Bolton2007}. This yields the optical depth, $\tau$,
for Lyman-$\alpha$ absorption as a function of velocity offset along
each line of sight, which can then be easily converted into a
transmitted flux fraction, $F=e^{-\tau}$, as a function of wavelength
or redshift.

\begin{figure*}
\centerline{\includegraphics[width=\linewidth]{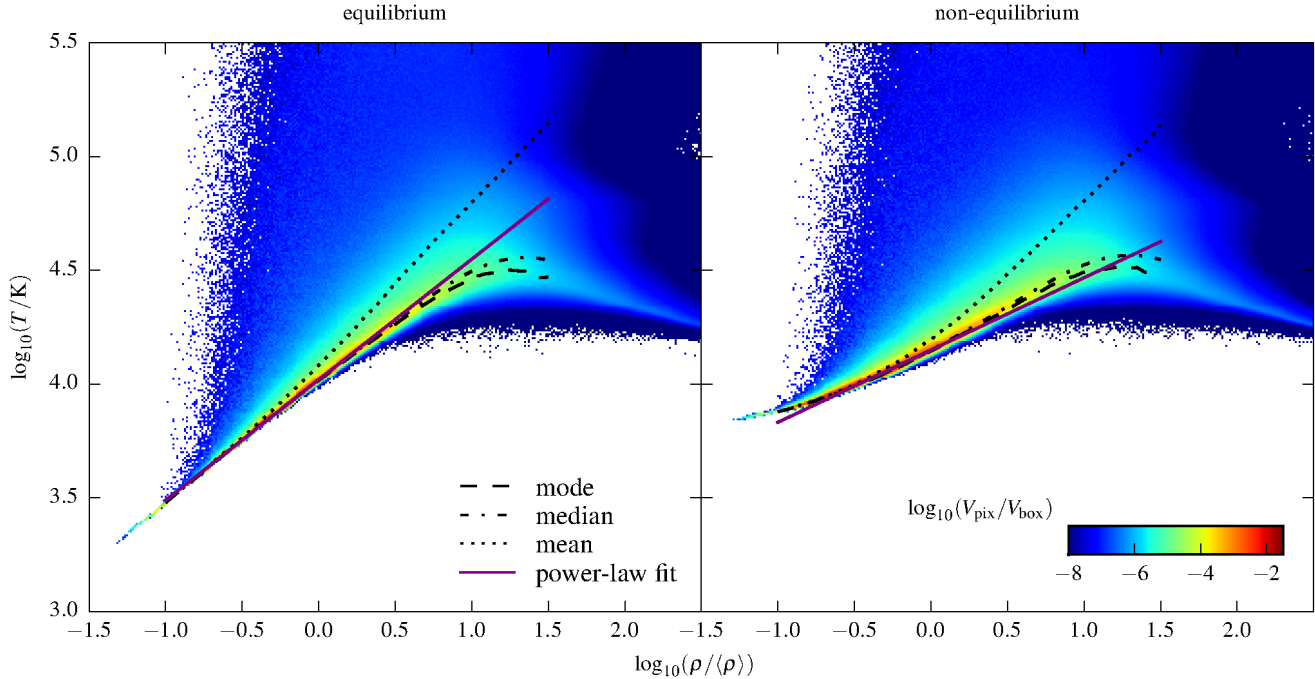}}
\caption{The volume-weighted distribution of gas in the
  temperature-density plane in equilibrium (\textit{left panel}, run
  L20N512 HM2012 eq.) and non-equilibrium (\textit{right panel}, run
  L20N512 HM2012 non-eq.) simulations with a HM2012 UVB. The
  results are given at $z=3.5$, i.e. shortly after the bulk of the He
  \textsc{ii} has been reionized in these models.}
\label{fig:rhoT}
\end{figure*}

\subsection{Measuring the temperature of the IGM}
\label{sec:temp_defs}

The thermal state of the IGM cannot be faithfully described by a single temperature. As has
been discussed many times a strong correlation between density and
temperature is expected due to the balance of photoheating/cooling and
adiabatic cooling/heating due to adiabatic expansion and compression
\citep[e.g.][]{Hui1997,Theuns1998,Valageas2002}. This is illustrated in 
Fig.~\ref{fig:rhoT}, where we display the volume-weighted
distribution of gas in the temperature-density plane from one of our
simulations. The results are shown for equilibrium and non-equilibrium ionization with 
a HM2012 UVB at $z=3.5$, i.e. shortly after
the bulk of the He\,\textsc{ii} has been ionized in these models (see
Fig.~\ref{fig:HI_HeII_fracs}).
A strong correlation between density and temperature is clearly
visible in the low-density IGM. It follows roughly a straight line in
this log-log plot. This has motivated many authors in the past to
approximate the temperature-density distribution by a power-law
relation $T = T_0 \Delta^{\gamma - 1}$ \citep[e.g.][]{Hui1997}, where
$T$ is temperature, $\Delta$ is the IGM density in units of the mean
cosmic baryon density and $T_0$ and $\gamma$ parametrize the
normalization and slope of the relation. Such power-law fits are
indicated by the \textit{purple solid} lines in Fig.~\ref{fig:rhoT}.

However, given the increased accuracy of recent observational
constraints and numerical predictions on the thermal state of the IGM,
it may no longer be justified to use a simple power-law to
characterize the temperature-density distribution, as this neglects
the width of the distribution and ignores deviations in the shape of
the relation. Such deviations can arise due to photoheating \citep[see
  e.g.][]{Furlanetto2008}. As shown in Fig.~\ref{fig:rhoT}, we find
them in particular in our non-equilibrium simulation shortly after
He\,\textsc{ii} reionization. There the logarithmic slope of the
relation increases with density (see a more detailed discussion about
this in Sec.~\ref{sec:T-rho}). Deviations from the power-law shape
also occur very prominently in models in which the IGM is heated by
TeV blazars \citep{Chang2012,Puchwein2012} as has been suggested by
\citet{Broderick2012}.

Fig.~\ref{fig:rhoT} also illustrates that even at fixed density
$\Delta$ the definition of an IGM temperature is ambiguous, as there
is a distribution of temperatures at each density. The width of this
distribution increases with increasing density (and also toward
  lower redshift) as more of the gas becomes shock heated. At
increasing density it therefore becomes more and more problematic to
neglect the width of the distribution by assuming a power-law
relation. For interpreting the Lyman-$\alpha$ forest, this becomes
more important at low redshifts as the forest is sensitive to higher
density at lower redshift. Furthermore, although not included in
our simulations, additional scatter in the temperature-density plane
from inhomogeneous heating during He\,\textsc{ii} reionization will
also blur the power-law relationship at low density
(e.g. \citealt{MeiksinTittley2012,Compostella2013}).  This will
further exacerbate the identification of a single power-law which
describes the IGM thermal state during and immediately following
He\,\textsc{ii} reionization.

In the remainder of the paper, we will thus distinguish the following
definitions of temperature at a given overdensity:
\begin{itemize}[leftmargin=5mm]

 \item $T_\textrm{mode}(\Delta)$ refers to the \textit{mode} of the
   distribution of the logarithms of the temperature at density
   $\Delta$. It corresponds to the temperatures at which the
   distributions shown in Fig.~\ref{fig:rhoT} attain the largest value
   along lines of constant density. Details about how
   $T_\textrm{mode}$ is computed from simulations are given in
   Appendix~\ref{sec:measure_eos}.

 \item $T_\textrm{median}(\Delta)$ refers to the \textit{median}
   temperature of all gas particles (of constant mass) with densities
   within 5\% of $\Delta$.

 \item $T_\textrm{mean}(\Delta)$ refers to the (mass-weighted)
   arithmetic \textit{mean} of the temperature of all gas particles
   with densities within 5\% of $\Delta$.
  
 \item $T_\textrm{power law}(\Delta)$ is computed by evaluating a
   power-law fit to the temperature-density relation at density
   $\Delta$. The power-law is defined by the points
   ($\Delta_\textrm{1}=10^{-0.5}$,$T_\textrm{median}(\Delta_\textrm{1})$)
   and
   ($\Delta_\textrm{2}=1$,$T_\textrm{median}(\Delta_\textrm{2})$). This
   is the same definition of $T(\Delta)$ that has been used in
   \citet{Becker2011}.

\end{itemize}

\noindent
Our different temperature definitions are indicated in
Fig.~\ref{fig:rhoT}. As expected from the increasing width of the
temperature-density distribution, these definitions differ more
strongly at higher density.

The IGM temperature directly affects the Lyman-$\alpha$ forest by the
thermal broadening of absorption lines. This makes it possible to
constrain the temperature based on Lyman-$\alpha$ absorption
spectra. However, the forest is not only sensitive to the temperature
at the time of absorption, i.e. the instantaneous temperature, but
also indirectly to the temperature at earlier times. The latter
affects the hydrodynamics of the IGM and thereby changes its density
distribution on small scales. In particular, the larger pressure at
higher temperature prevents the collapse of gas into small
structures. This is referred to by the term \textit{Jeans smoothing}
\citep{Gnedin1998,Hui1999,Theuns2000}.

In Sec.~\ref{sec:results}, we will also compare our simulation
predictions of the IGM temperature to observational constraints from
\citet{Becker2011}. In the following, we will briefly summarize how
their method to constrain the temperature works. The main steps are:
\begin{itemize}[leftmargin=5mm]

 \item A set of cosmological hydrodynamical reference simulations with
   different (rescaled) photoheating rates are performed. Due to the
   different assumptions for the photoheating rates, the simulations
   span a range of IGM temperatures at each redshift.

 \item Synthetic Lyman-$\alpha$ forest spectra are computed from the
   reference simulations. The curvature of the spectra (defined by
   Eq.~(\ref{eq:def_kappa})) are computed.

 \item Temperatures $T_\textrm{power law}(\Delta)$ are computed from
   the reference simulations. At each redshift, the density
   $\bar{\Delta}(z)$ at which the tightest relation between spectral
   curvature and $T_\textrm{power law}(\Delta(z))$ occurs is
   identified. The spectral curvature is then considered as a proxy
   for $T_\textrm{power law}(\bar{\Delta}(z))$.

 \item The spectral curvature is measured from the observed
   Lyman-$\alpha$ spectra. It is then converted to a temperature
   constraint at density $\bar{\Delta}(z)$ using the
   curvature-$T_\textrm{power law}(\bar{\Delta}(z))$ relation that was
   obtained from the reference simulations.
\end{itemize}

\noindent
In practice, this method is not only sensitive to the instantaneous
temperature but also to the previous thermal history of the IGM, as
both Doppler broadening and Jeans smoothing affect the curvature of
the spectra. This was already pointed out in \citet{Becker2011} and
will be further discussed in Sec.~\ref{sec:curv_temp}.

\begin{figure*}
\centerline{\includegraphics[width=\linewidth]{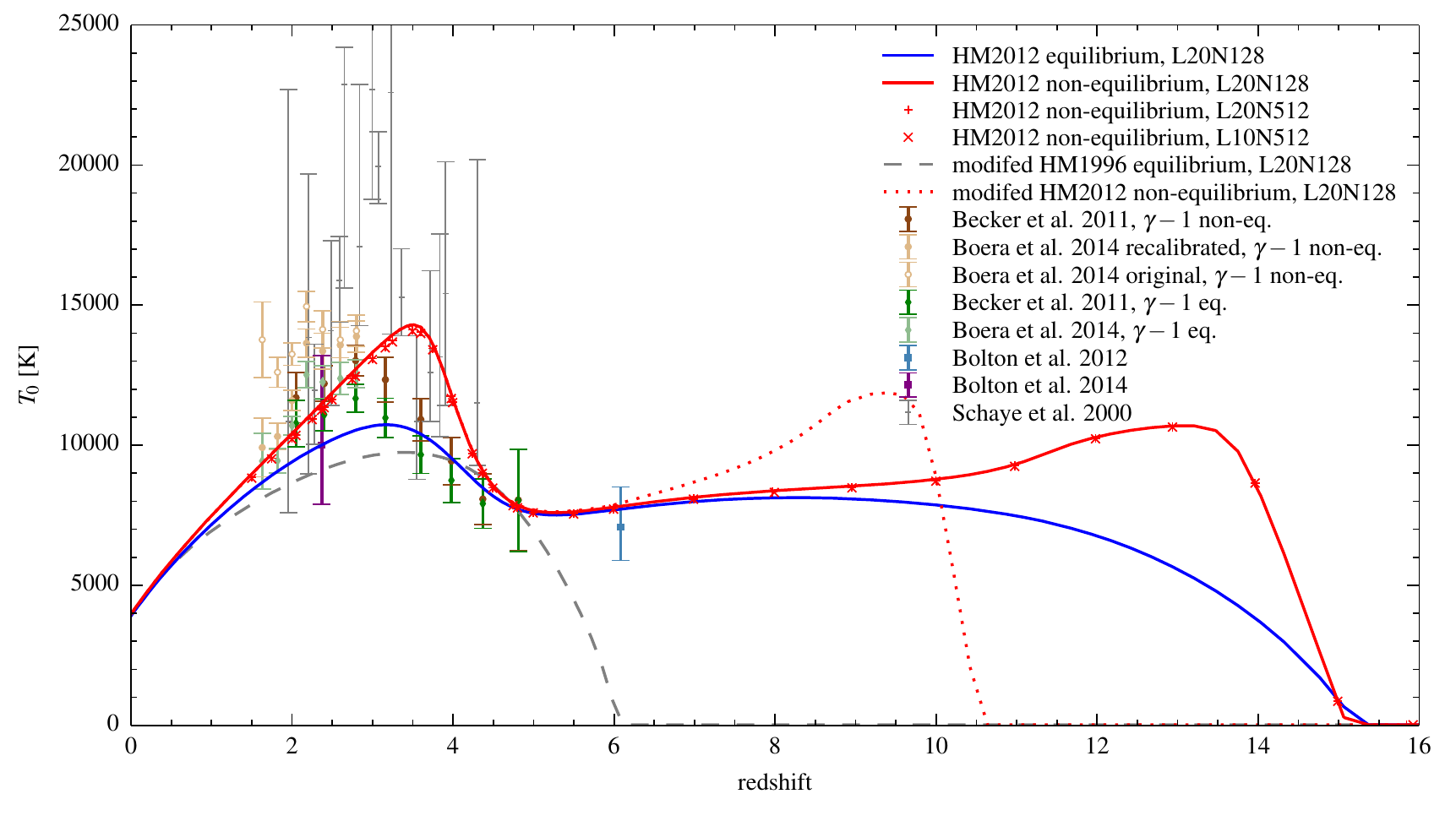}}
\caption{The IGM temperature at mean density ($T_\textrm{median}(\Delta=1)$) as a function of
  redshift. Results for equilibrium and non-equilibrium simulations
  with the HM2012 UVB are shown. Almost the same
  temperatures are found for runs with different numerical
  resolutions. Also indicated are the temperatures in an equilibrium
  run with a modified HM1996 background, as well as the temperatures
  in a non-equilibrium run with a modified HM2012 background (see
  Sec.~\ref{sec:curv_temp} for the latter). Observational constraints
  from \citet{Schaye2000}, \citet{Becker2011},
  \citet{Bolton2012,Bolton2014} and \citet{Boera2014} are shown for
  comparison.}
\label{fig:T0}
\end{figure*}

\section{Results}
\label{sec:results}
\subsection{The thermal history of the IGM}
\label{sec:thermal_history}

\subsubsection{The temperature at mean density}
\label{sec:T0}

In Figure~\ref{fig:T0}, we compare temperature predictions using the
the HM2012 UVB to observational constraints. To
facilitate comparison with the literature, we do this first at mean
density. Note, however, that depending on redshift the Lyman-$\alpha$
forest may be more sensitive to other densities.

Before we discuss the level of agreement with observations, however,
it is worthwhile to briefly consider the differences between
equilibrium and non-equilibrium simulations. It can be clearly seen
that the non-equilibrium treatment results in a much larger
temperature increase during the almost simultaneous H\,\textsc{i} and
He\,{\sc i} reionization between redshifts $\sim 15$ and $\sim 12$, as
well as during He\,\textsc{ii} reionization between redshifts $\sim 5$
and $\sim 3.5$. In the equilibrium treatment, an increase in the
photoionization rates results in an unrealistic instantaneous increase
of the ionized fractions as they are directly set to the new
equilibrium value. However, as the photoheating rates are computed
from the abundance of neutral or singly-ionized atoms, this
underestimates the photoheating rates (see Appendix~\ref{sec:temp_ne}
for a more detailed discussion of this). The significant differences
in the temperature re-emphasizes the potential importance of
accounting for out-of-equilibrium ionized fractions in cosmological hydrodynamical simulations.

Figure~\ref{fig:T0} displays a number of observational constraints
that are based on the observed curvature of the Lyman-$\alpha$ forest
transmitted flux \citep{Becker2011,Boera2014} and the observed
Lyman-$\alpha$ absorption line widths
\citep{Schaye2000,Bolton2012,Bolton2014}. Our simulation predictions
are in excellent agreement with the \citet{Bolton2012,Bolton2014}
constraints, and also in reasonably good agreement with the
\citet{Becker2011} and \citet{Boera2014} measurements. Both of these
latter studies quote values for the overdensity\footnote{For the purpose of direct comparison in this work, the
    \cite{Boera2014} measurements have been recalibrated assuming the
    same \lya effective optical depth, $\tau_{\rm eff}(z)$, as
    \cite{Becker2011}.  This increases the characteristic
    densities reported by \cite{Boera2014} by 20-40 per cent (E. Boera,
    private communication). Fig.~\ref{fig:T0} displays both the original and rescaled \cite{Boera2014} constraints.} to which their measurements are
sensitive. We can thus scale the temperature to the mean density using
the slope of the temperature-density relation from our
simulations. The rescaled values are shown both using the slope at
mean density from our equilibrium and our non-equilibrium runs (see
Sect.~\ref{sec:T-rho}, and Appendix~\ref{sec:measure_eos} for how the
slope is measured).

In the non-equilibrium case, the simulation predictions agree
well with the \citet{Schaye2000} measurements, except for the two
data points at $z\sim 3$.  We note, however, that these data were
obtained using \textsc{hydra} hydrodynamical simulations
(\citealt{Couchman1995}) which use outdated cosmological parameters
and have a low dynamic range by present day standards.  It is thus
somewhat unclear how instructive the observed level of agreement
is. The non-equilibrium simulations also deviate somewhat from the
\citet{Becker2011} constraints in the redshift range $3 < z < 4.5$. In
particular, the models predict a slightly earlier and larger increase
in the gas temperature.  We will explore the cause of this deviation
in Sec.~\ref{sec:deviations}. Surprisingly, the equilibrium runs agree
better with the data at these redshifts. We will, however, show in
Sec.~\ref{sec:deviations} that this better agreement is largely a
coincidence. Furthermore, the equilibrium runs underpredict the
temperature more strongly at $z < 3$.  Lastly, note also that the
exact location and height of the temperature peak corresponding to
He\,\textsc{ii} reionization does not only depend on the UV
background, but at some level also on the other adopted rate
coefficients in the simulations \citep[see e.g.][]{Iliev2006,Lukic2014}.

The overall good agreement between IGM temperatures obtained by
simulations with a HM2012 UVB and observational constraints
is very reassuring. This suggests that when using this UVB
model a rescaling of the photoheating rates that has routinely been
employed in the past to obtain IGM temperatures consistent with
observational data \citep[e.g.][]{Viel2004,Jena2005} may no longer
be required. Indeed, the lower temperatures inferred in recent
observational studies, coupled with improved constraints on the
underlying cosmology and significantly higher resolution
simulations, may account for most of the discrepancy between the
observed and simulated velocity widths of \lya forest absorption
lines noted in the early analyses in this field
(e.g. \citealt{Bryan99,Theuns1999,Meiksin2001}).

\begin{figure}
\centerline{\includegraphics[width=\linewidth]{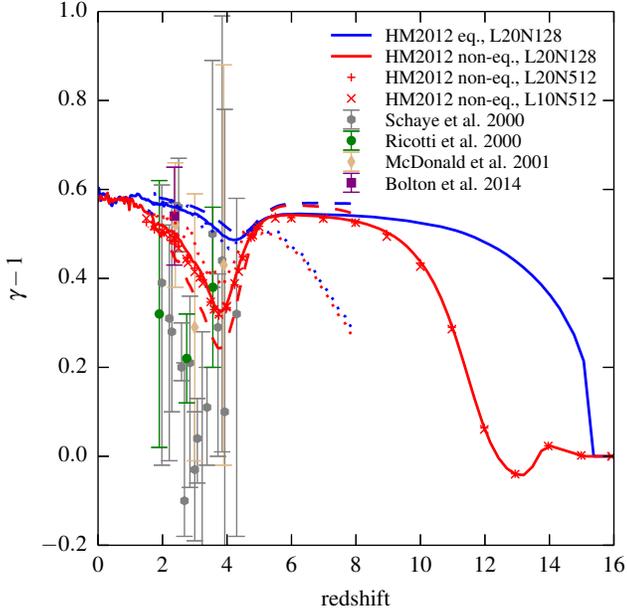}}
\caption{Logarithmic slope $\gamma - 1$ of the temperature-density
  relation at mean density as a function of redshift. Results for
  equilibrium and non-equilibrium simulations with the HM2012 UVB
  are shown (\textit{blue} and \textit{red solid}
  lines). Almost the same slopes are found for runs with different
  numerical resolutions. Observational constraints from
  \citet{Schaye2000}, \citet{Ricotti2000}, \citet{McDonald2001} and
  \citet{Bolton2014} are shown for comparison. The \textit{dashed} and
  \textit{dotted} lines show the slopes of the simulated
  temperature-density relations at $\Delta = 10^{-0.5}$ and
  $10^{0.5}$, respectively.}
\label{fig:gamma_minus_1}
\end{figure}

\subsubsection{The slope of the temperature-density relation}
\label{sec:T-rho}

We now also investigate how the slope of the temperature-density
relation in our simulations compares to observational constraints, and
discuss differences between the equilibrium and the non-equilibrium
simulations. Figure~\ref{fig:gamma_minus_1} shows the logarithmic
slope $\gamma - 1$ of the temperature-density relation at mean
density, as well as at the densities $\Delta = 10^{-0.5}$ and
$10^{0.5}$. Appendix~\ref{sec:measure_eos} explains how these slopes
were measured from the simulations.

During He\,\textsc{ii} reionization, the IGM is photoheated everywhere
by the same spectrum in our simulations with a homogeneous UVB. We thus
initially expect a roughly constant temperature increase, independent
of density.  At higher initial (i.e. before He\,\textsc{ii}
reionization) temperature this corresponds to a lower increase in the
logarithm of the temperature, so that the temperature-density relation
flattens in log-log space. We, indeed, observe this behaviour in our
non-equilibrium simulations. In the equilibrium run, however, only a
small reduction in the logarithmic slope is found. This is a
consequence of the much lower temperature boost in this run (see
Fig.~\ref{fig:T0}). Furthermore, the heating is always density
dependent in the equilibrium computation, as the equilibrium ionized
fraction on which the heating rate is based depends on the
recombination rate and, thus, on density. The consequences of this are
discussed in full detail in Appendix~\ref{sec:slope_ne}. The main
effect is that the temperature-density relation retains its power-law
shape in the equilibrium treatment with only a slight reduction of the
slope. This can be seen in the volume-weighted temperature-density
phase-space plot shown in the \textit{left} panel of
Fig.~\ref{fig:rhoT}. It shows results for $z=3.5$, i.e. right after
the bulk of the He\,\textsc{ii} has been reionized. The \textit{right}
panel of Fig.~\ref{fig:rhoT} shows the stronger flattening of the
relation in the non-equilibrium treatment. The flattening is also, as
expected, more pronounced at low density, i.e. for lower initial
temperature, and causes deviations of the temperature-density relation
from the power-law shape. The same effects are illustrated by the
\textit{dashed} and \textit{dotted} lines in
Fig.~\ref{fig:gamma_minus_1}, which show the logarithmic slope $\gamma
- 1$ of the temperature-density relation at densities $\Delta =
10^{-0.5}$ and $10^{0.5}$, respectively.

Our non-equilibrium simulations are overall in good agreement with the
constraints from \citet{Ricotti2000}, \citet{McDonald2001} and
\citet{Bolton2014}. The error bars are, however, admittedly large. The
\citet{Schaye2000} data favours a flatter or even inverted
temperature-density relation around $z \approx 3$, although again
the quoted uncertainties are large and the dynamic range of the
simulations in the earlier studies is low by present day standards.
As for the temperature, the agreement with the constraints on the
slope may improve if He\,\textsc{ii} reionization happened
slightly later in our simulations.

\subsection{H I and He II Lyman-$\alpha$ effective optical depths}
\label{sec:tau}

\begin{figure*}
\centerline{\includegraphics[width=\linewidth]{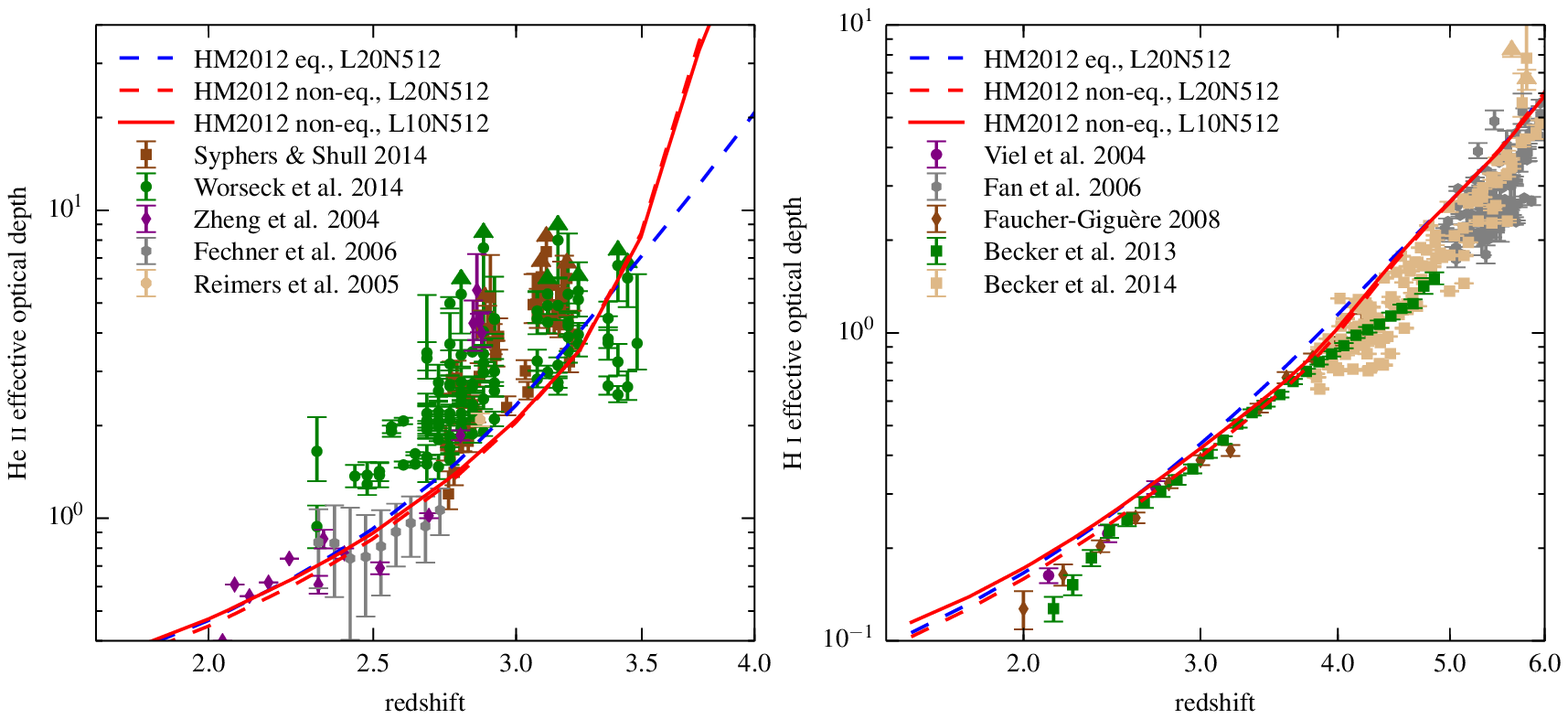}}
\caption{Effective optical depths for He\,\textsc{ii} ({\it left
    panel}) and H\,\textsc{i} ({\it right panel}) Lyman-$\alpha$
  absorption as a function of redshift. The results are based on
  equilibrium and non-equilibrium simulations with the HM2012 UV
  background. Observational constraints on the He\,\textsc{ii}
  effective optical depth from \citet{Zheng2004}, \citet{Reimers2005},
  \citet{Fechner2006}, \citet{Syphers2014} and \citet{Worseck2014} are
  shown for reference. For the H\,\textsc{i} effective optical depth,
  we compare to data from \citet{Viel2004}, \citet{Fan2006},
  \citet{Faucher-Giguere2008}, \citet{Becker2013} and
  \citet{Becker2014}. The x-axis is linear in $\log(1+z)$ so that a
  power-law evolution corresponds to a straight line.}
\label{fig:tau}
\end{figure*}

We now turn to analysing the \lya forest in our hydrodynamical
simulations. Using the methods outlined in Sec.~\ref{sec:spectra},
we have computed synthetic H\,\textsc{i} and He\,\textsc{ii}
Lyman-$\alpha$ absorption spectra for our equilibrium and
non-equilibrium simulations. The effective optical depths $\tau_{\rm
  eff} = - \ln(\langle F \rangle)$ in our equilibrium and
non-equilibrium simulations with a HM2012 UVB are compared to
data in Fig.~\ref{fig:tau}. Here $\langle F \rangle$ is the mean
transmitted fraction obtained after continuum removal. 

The left panel of Fig.~\ref{fig:tau} displays the
He\,\textsc{ii} Lyman-$\alpha$ effective optical depths that we obtain
with a spatially homogeneous HM2012 UVB. These appear to be in
good agreement with the slowly rising opacity measurements obtained
with FUSE and HST at $z \lesssim 2.8$ \citep{Zheng2004,Fechner2006,
  Worseck2014,Syphers2014}.  At $z>2.8$ the opacity measurements rise
more rapidly and spatial fluctuations appear to strongly increase as
expected at the tail-end of He\,\textsc{ii} reionization
(although see \cite{Davies2014}; \cite{Khaire2013}). The He\,\textsc{ii} opacity
data for $z \lesssim 3.3$ thereby suggests that He\,\textsc{ii}
reionization occurs too early in our simulations, consistent with the
apparently too early temperature increase found in
Sec.~\ref{sec:T0}. Somewhat surprisingly, however, at $z>3.3$
\citet{Worseck2014} have recently measured significantly lower optical
depths than our simulations predict. Although the statistical
significance of these data may still be relatively low
(\citealt{Compostella2014}), our simulations nevertheless suggest
that it may be difficult to reconcile the timing of the temperature
increase as measured by \citet{Becker2011} with the slow evolution of
He\,\textsc{ii} opacities found in \citet{Worseck2014}.  If both
measurements are confirmed by future studies, this may indicate that
it is not (only) the photoheating of He\,\textsc{ii} that is
responsible for the observed temperature increase. We will discuss the
timing of He\,\textsc{ii} reionization in more detail in the next
section.

In the right panel of Fig.~\ref{fig:tau} we compare the H\,\textsc{i}
opacity in our simulations with observation. The overall agreement is
good, but there are also significant differences. As already noted by
\citet{BeckerBolton2013}, the HM2012 UVB model appears to
significantly under predict the photoionization rate at $z>4$. This
results in simulated opacities which are too large at these
redshifts. Similarly to what is seen for He\,\textsc{ii} at $z \gtrsim
2.8$, fluctuations in the H\,\textsc{i} effective optical depth start
to increase rapidly at $z>5.5$, presumably due to the large spatial
fluctuations in the UVB flux expected at the tail end of
reionization in the immediate aftermath of the percolation of ionized
regions \citep{Miralda-Escude2000,WyitheLoeb2006,Fan2006,Becker2014}.
The H\,\textsc{i} Lyman-$\alpha$ effective optical depth is in good
agreement with the data in the range $2.5 \lesssim z \lesssim 4$. As
noted recently by \citet{Kollmeier2014}, the photoionization rate in
the HM2012 model appears to be too low to reproduce the column-density
distribution of the low-redshift Lyman-$\alpha$ forest at $z\sim
0.1$. We, thus, expect our simulations to overpredict the
H\,\textsc{i} opacity at low-redshift. This is indeed observed for
$z<2.5$.

Lastly, comparing the equilibrium and non-equilibrium results, we note
that in the latter run the larger temperatures due to the more
efficient He\,\textsc{ii} photoheating translate to somewhat smaller
effective optical depths in the H\,\textsc{i} Lyman-$\alpha$
forest. However, as this effect is present over a large range in
redshift, roughly $2 \lesssim z \lesssim 4.5$, no sharp features are
predicted in the redshift evolution of the H\,\textsc{i}
Lyman-$\alpha$ effective optical depth (cf. \citealt{Theuns2002b,Bernardi2003,DallAglio2008,Faucher-Giguere2008b}). In particular, there is
no evidence for a dip in the effective optical depth evolution
associated with He\,\textsc{ii} reionization (see also
\citealt{Bolton2009b}).

\subsection{What causes the remaining discrepancies between simulations and observations?}
\label{sec:deviations}

In the analysis above we have demonstrated that the predicted IGM
temperatures are overall in good agreement with observations. We
shall discuss further why this is the case in
Sec.~\ref{sec:discussion}.  Some deviations from the
\citet{Becker2011} constraints were, however, found during
He\,\textsc{ii} reionization. Furthermore, the predicted
He\,\textsc{ii} effective optical depths are somewhat different than
observed. We will investigate two possible causes for these
deviations: (1) The curvature method used by \citet{Becker2011} may not be able to reproduce
sharp peaks in the temperature evolution well. (2) He\,\textsc{ii}
reionization may not happen at the correct time in our simulations.

\subsubsection{Comparing directly to the temperature measurements obtained with the curvature method}
\label{sec:curv_temp}

In Fig.~\ref{fig:T0}, we scaled the \citet{Becker2011} temperature
constraints to the mean density. Here, we perform a more direct
comparison, i.e. we apply the method that \citet{Becker2011} used to
constrain the IGM temperature to our simulations.  More precisely, we
compute synthetic Lyman-$\alpha$ forest spectra from our simulations
as detailed in Sec.~\ref{sec:spectra}. For the results presented in
this subsection, we rescale the optical depths such that the mean
transmission is in agreement with the values measured by
\citet{Becker2013}. We then apply the curvature-temperature method as
described in Sec.~\ref{sec:temp_defs} to the synthetic spectra. That
is, we calibrate a relation between spectral curvature and IGM
temperature at the density $\bar{\Delta}(z)$ at which it was measured
by \citet{Becker2011} (see their Table~3) using exactly the same
reference simulations they employed.  Finally, we use this relation to
measure the IGM temperature in our new simulations. This procedure
allows us to directly compare our simulated spectral
curvature-temperatures to the observational constraints.

\begin{figure}
\centerline{\includegraphics[width=\linewidth]{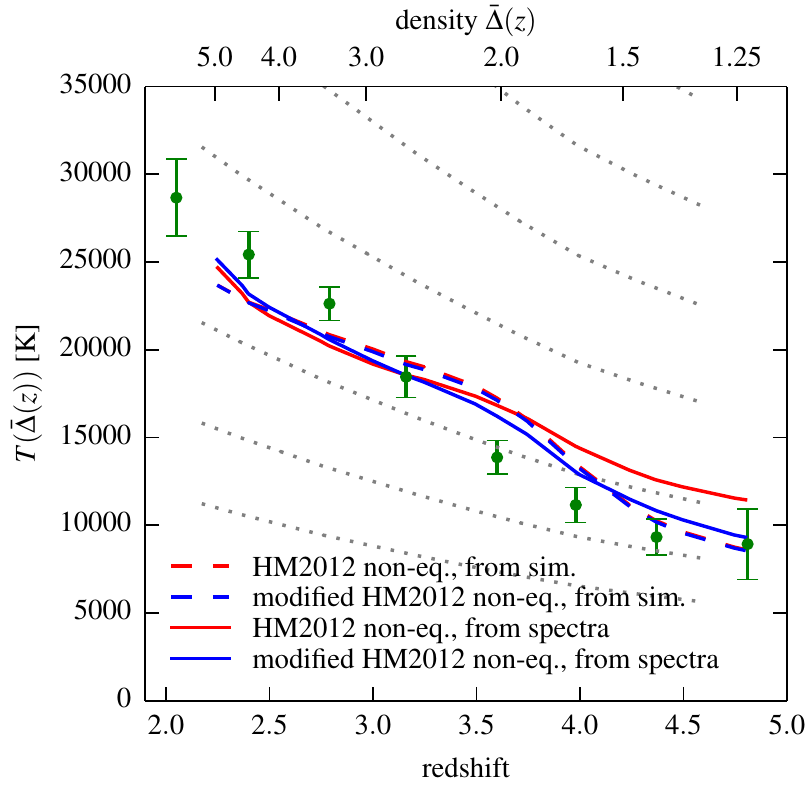}}
\caption{Comparison of IGM temperatures computed directly from the
  simulations (\textit{dashed}, indicating $T_\textrm{power law}$ in
  the L10N512 runs) and obtained from the curvature of synthetic
  Lyman-$\alpha$ forest spectra (\textit{solid}). The temperatures are
  shown at the densities $\bar{\Delta} (z)$ to which the
  curvature method is most sensitive. Results for a non-equilibrium
  simulation with the HM2012 UVB are shown by \textit{red}
  lines. The \textit{blue} lines indicate results for a
  non-equilibrium run with the modified HM2012 background in which
  H\,\textsc{i} and He\,\textsc{i} reionization happen later. The
  \citet{Becker2011} observational constraints (\textit{green circles} and \textit{error bars}) are shown for
  reference. The temperatures in the reference simulations, which
  these authors used to calibrate a curvature-temperature relation,
  are indicated by the \textit{grey dotted} lines.}
\label{fig:curv_method}
\end{figure}

This is illustrated in Fig.~\ref{fig:curv_method}. The \textit{red
  dashed} line indicates the temperature $T_\textrm{power
  law}(\bar{\Delta}(z))$ in our non-equilibrium simulation with a
HM2012 UVB. The \textit{red solid} line shows the temperature
that is estimated from the synthetic spectra using the curvature
method. At redshifts larger than $\sim 4$ the curvature method
overestimates the temperature in our simulation, while it is slightly
underestimated in the range $2.5 \lesssim z \lesssim 3.7$.

A difference in the amount of Jeans smoothing in our simulation
compared to the reference simulations used in \citet{Becker2011} could
cause such a discrepancy, as it will change the spectral curvature
even for identical instantaneous temperatures. In order to assess
whether this can indeed explain the deviations, we have performed a
non-equilibrium simulation with a modified HM2012 background. The
modification was chosen such that the instantaneous temperature below
redshift 6 is unchanged, while shifting H\,\textsc{i} and
He\,\textsc{i} reionization to a lower redshift, $z\sim 10$, thereby
reducing the amount of Jeans smoothing. This brings our simulation
much closer to the reionization redshift, $z \sim 9$, used in the
\citet{Becker2011} reference runs and should, thus, result in a more
similar Jeans smoothing. 

As shown in Fig.~\ref{fig:curv_method}, the discrepancy between the
temperature in the simulation and measured from synthetic spectra at
$z>4$ is alleviated with this modified UVB, albeit some
difference remains. This indicates that the discrepancy was indeed at
least partially caused by a difference in Jeans smoothing and
highlights that the curvature method is not only sensitive to
instantaneous temperature, but to a combination of instantaneous
temperature and Jeans smoothing, as has already been discussed in
\citet{Becker2011}.

To understand this degeneracy better, we have investigated to which
spatial scales the spectral curvature is most sensitive and to what
extent they are affected by Jeans smoothing. Full details are given in
Appendix~\ref{sec:curv_vs_powspec}. Our main finding is that the
contribution of a specific scale to the mean square of the curvature
$\kappa$ is roughly given by $\textrm{d} \langle \kappa^2 \rangle /
\textrm{d} (\ln k) \propto k^5 P(k)$, where $k$ is is the wavenumber
corresponding to that scale and $P(k)$ is the flux power
spectrum. Most of the contribution comes indeed from scales that are
too large to be fully dominated by the thermal cutoff, so that Jeans
smoothing also plays a significant role. This suggests that for
spectra with sufficiently high resolution, it might be favourable to
apply a window function to the power spectrum that gives more weight
to smaller scales to get a more accurate proxy of instantaneous
temperature. Metal contamination may, however, be a more severe
problem there.

At redshifts $2.5 \lesssim z \lesssim 3.7$, the curvature method
somewhat underpredicts the simulated temperature, both for the
original and the modified HM2012 background. This is most likely also
caused by differences in the Jeans smoothing compared to the reference
simulations. In particular, the reference runs used in
\citet{Becker2011} have a fairly smooth thermal evolution, as shown by
the \textit{grey dotted} lines in Fig.~\ref{fig:curv_method}. Our runs
exhibit instead a significant heating due to He\,\textsc{ii}
reionization. Thus, after He\,\textsc{ii} has been reionized, we
effectively compare the curvatures to a reference model which has much
higher temperature before He\,\textsc{ii} reionization and, thus, more
Jeans smoothing. This biases the curvature temperatures slightly low.

However, comparing the curvature temperatures from our synthetic
spectra to the observational constraints, we still find that the
temperature increase due to He\,\textsc{ii} reionization happens
somewhat too early in our simulations even if we account for possible
different amounts of Jeans smoothing. This will be discussed in more
detail in Sec.~\ref{sec:min_ion_model}. At low redshift, i.e. $z
\lesssim 3$, the observed curvature temperature is larger than the
value computed from the simulations. Thus, additional heating may be
required there. This suggests that either photoheating may be more
efficient than in our non-equilibrium simulations, or perhaps that
there is another source of IGM heating like TeV blazars
\citep{Broderick2012}. It was shown by \citet{Puchwein2012} and
\citet{Boera2014} that the latter could be responsible for this.

\subsubsection{Did He\,II reionization happen somewhat later?}
\label{sec:min_ion_model}

The temperatures and ionization fractions which we have shown so far
were based on the photoionization and photoheating rates presented in
HM2012, i.e. in their Table~3 (the only exceptions being the results
based on the modified HM1996 and the modified HM2012 backgrounds). The
comparison to the \citet{Becker2011} constraints indicates that
He\,{\sc ii} reionization may happen somewhat too early in our
simulations. The same conclusion was found by comparing the effective
optical depth of the He\,{\sc ii} Lyman-$\alpha$ forest to
observational constraints, although interestingly there are some
recent measurements from \citet{Worseck2014} at $z \gtrsim 3.3$ which
contradict this. 

It is however, not obvious that our simulations do actually predict a
He\,{\sc ii} reionization history consistent with the evolution of the
ionizing emissivity, $\epsilon_{\nu}$, assumed in HM2012; applying the
HM2012 photoionization and heating rates as a spatially uniform
UVB model does not account
properly for the consumption of photons in overdense regions due to
recombinations. This effectively assumes the mean free path of He\,{\sc ii} ionizing photons is much greater than the size of the simulation box, which is not true during reionization. Furthermore, the ionized fractions in our simulations are inconsistent with what one would expect based on the ionizing emissivity and the estimated number of recombinations as also calculated in HM2012 in their \textit{``Minimal reionization
model''}. This suggests that either the mean free path,
$\lambda_{\nu}$, adopted in the HM2012 calculation to convert emissivity to photoionization rate, i.e. $\Gamma_{\rm i} \propto \epsilon_{\nu}\lambda_{\nu}$, differs from the mean free path in our simulations or that the number of recombinations is incompatible.
In the following we try to account and correct for this in an approximate
manner by using the volume filling factor of He\,{\sc iii} regions
computed in the ``Minimal reionization model'' in HM2012.

Following HM2012 we assume that the ionized, i.e. He\,{\sc iii},
volume fraction $Q_{\textrm{He} \, \textsc{iii}}$ evolves according to
\begin{equation}
 \frac{\textrm{d} Q_{\textrm{He} \, \textsc{iii}}}{\textrm{d} t} = \frac{\dot{n}_{\textrm{He} \, \textsc{iii}, \textrm{ion}}}{\langle n_{\textrm{He}} \rangle} - \frac{Q_{\textrm{He} \, \textsc{iii}}}{\langle t_{\textrm{He} \, \textsc{iii}, \textrm{rec}} \rangle},
\end{equation}
where $t$ is time, $\dot{n}_{\textrm{He} \, \textsc{iii},
  \textrm{ion}}$ is the production rate of He\,{\sc ii} ionizing
photons per unit volume, $\langle n_{\textrm{He}} \rangle$ is the mean
He\,{\sc ii} number density and $\langle t_{\textrm{He} \,
  \textsc{iii}, \textrm{rec}} \rangle$ is the mean He\,{\sc iii}
recombination time which is based on a clumping factor
$C_{\textrm{IGM}} = 1 + 43 z^{-1.71}$ obtained from simulations by
\citet{Pawlik2009}. This model, hence, explicitly accounts for the
production and consumption of ionizing photons.

\begin{figure}
\centerline{\includegraphics[width=\linewidth]{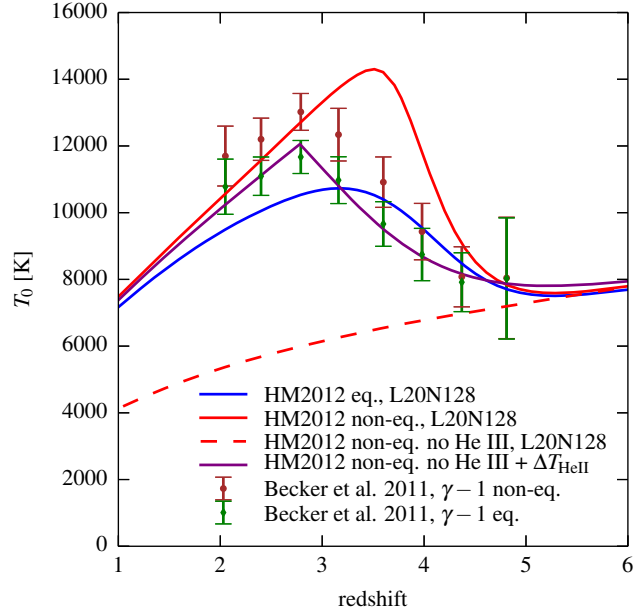}}
\caption{The IGM temperature at mean density ($T_\textrm{median}(\Delta=1)$) as a function of
  redshift. The \textit{blue solid} and \textit{red solid} curves, as
  well as the \citet{Becker2011} observational constraints are the same as in
  Fig.~\ref{fig:T0}. Additionally, the \textit{red dashed} curve shows
  a simulation in which the ionization of He\,\textsc{ii} and
  associated heating was turned off. Based on the latter simulation,
  the HM2012 He\,\textsc{iii} volume filling factor and the excess
  energy per He\,\textsc{ii} ionization, we compute an estimate of the
  temperature at mean density (\textit{purple}), which is in very good
  agreement with the observations.}
\label{fig:T0_zoom}
\end{figure}

This model results in a somewhat later reionization of He\,{\sc
  ii}. We correct the thermal evolution to account for this in the
following way. We start with a non-equilibrium simulation in which
He\,{\sc ii} ionization is turned off, i.e. for which the He\,{\sc ii}
photoionization and photoheating rates are set to zero. The thermal
evolution of the IGM in this simulation $T_{0,\textrm{no HeIII}}(z)$ is indicated in
Fig.~\ref{fig:T0_zoom} by the \textit{red dashed} curve. Next, we
modify this temperature by the following procedure.  Starting
before He\,{\sc ii} reionization, we compute for each timestep the
change in $Q_{\textrm{He} \, \textsc{iii}}$ implied by the HM2012
``Minimal reionization model''.\footnote{This was achieved by
  interpolating a table of the volume filling factors that is provided
  by Francesco Haardt and Piero Madau:
  \url{http://www.ucolick.org/~pmadau/CUBA/Media/Q.out}} We then
compute the change in the average temperature at mean density by assuming that
the newly reionized volume fraction was heated by a temperature
$\Delta T_{\textrm{He}\, \textsc{ii} \, \textrm{reion}}(z)$. The latter is computed using the excess energy
per He\,{\sc ii} reionization implied by the HM2012 model, i.e. using
the ratio of photoheating to photoionization rates, and accounting for
the change in the particle number due to He\,{\sc ii}
reionization. Finally, we integrate the temperature changes to get the
overall increase $\Delta T_{0}(z)$
of the average temperature at mean density due to He\,{\sc ii}
reionization. This is done according to
\begin{align}
 \Delta T_{0}&(z_{i+1}) = \, \Delta T_{0}(z_{i}) \, \left( \frac{1+z_{i+1}}{1+z_i} \right)^2 \nonumber \\
                               & + \left[Q_{\textrm{He} \, \textsc{iii}}(z_{i+1}) - Q_{\textrm{He} \, \textsc{iii}}(z_i) \right] \Delta T_{\textrm{He}\, \textsc{ii} \, \textrm{reion}}(z_i) \nonumber \\
                               & + Q_{\textrm{He} \, \textsc{iii}} \left[\Delta T_{\textrm{heat-cool},\textrm{He} \, \textsc{iii}} - \Delta T_{\textrm{heat-cool},\textrm{He} \, \textsc{ii}}\right],
\label{eq:temp_heii_reion}
\end{align}
for the timestep from redshift $z_i$ to $z_{i+1}$. The first term on
the right-hand side accounts for adiabatic cooling due to the Hubble
expansion. The second term on the right-hand side describes the
heating by He\,{\sc ii} reionization. The third term accounts for the
difference in heating and cooling between He\,\textsc{ii} and He\,\textsc{iii}
regions at fixed volume fraction. Full details how this term and $\Delta T_{\textrm{He}\, \textsc{ii} \, \textrm{reion}}$ are computed are given in Appendix~\ref{sec:ion_history_to_T}. Also note that Eq.~(\ref{eq:temp_heii_reion}) assumes that the fraction of mean density regions in which He\,\textsc{ii} has already been ionized traces the volume filling factor $Q_{\textrm{He} \, \textsc{iii}}$. In reality small deviations might exist but they are unlikely to be larger than other uncertainties in the HM2012 ``Minimal reionization
model''. The \textit{purple} curve in
Fig.~\ref{fig:T0_zoom} shows the sum of the temperature obtained in
the run without He\,{\sc ii} reionization and the average heat boost due to He\,{\sc ii} reionization, i.e. $T_0(z) = T_{0,\textrm{no HeIII}}(z) + \Delta T_{0}(z)$. It thus shows an estimate of the average IGM temperature at mean density that is based on the HM2012 He\,\textsc{iii} volume
filling factor, which is computed with their ``Minimal reionization
model'', and excess energy per He\,\textsc{ii} ionization.

As can be clearly seen, the later reionization of He\,{\sc ii} in this
model results in a temperature evolution that is in remarkably good
agreement with the \citet{Becker2011} constraints. This illustrates
that the thermal evolution of the IGM between redshifts 5 and 2.5 is
very sensitive to when He\,{\sc ii} is reionized. It also suggests
that the He\,\textsc{iii} volume filling factor as estimated in HM2012
is broadly consistent with the observed thermal history.
It is also worth noting that in
the light of these findings the better agreement of the equilibrium
run (compared to the non-equilibrium run) with the \citet{Becker2011}
temperature constraints in the redshift range $3 < z < 4.5$ (as shown
in Fig.~\ref{fig:T0}) appears to be a coincidence. In particular, the
artificial delay between reionization and photoheating that is present
in the equilibrium run mimics a later reionization of He\,\textsc{ii}.

\section{Discussion}
\label{sec:discussion}

Should our cosmological hydrodynamical simulations with a homogeneous UVB reproduce the
temperature measurements of the IGM at $2<z<5$ as well as they do?
There are some
aspects of the problem this calculation will not capture adequately.
However, as we have demonstrated in Sec.~\ref{sec:hmuvb}, although
the HM2012 rates are applied as a homogeneous UVB in our
simulations, they do self-consistently follow the
transition from optically thick to thin heating. Nevertheless, it is not
entirely clear how well this captures the volume average of the inhomogeneous reionization process.
When incorporating non-equilibrium effects,
as shown in Fig.~\ref{fig:T0} and Fig.~\ref{fig:gamma_minus_1}, the
HM2012 model predicts a temperature increase for a gas parcel at
mean density of $\Delta T\approx7000\,K$ and $\gamma-1 \sim
0.3$ following He\,\textsc{ii} reionization.  In
comparison\footnote{Note that both HM2012 and the radiative transfer
simulations of \cite{McQuinn2009} and \cite{Compostella2013}
assume He\,\textsc{ii} reionization is driven by quasars with UV
spectra $L_{\nu}\propto \nu^{-1.6}$ (\citealt{Telfer2002}).}
radiative transfer simulations typically find a somewhat larger
average boost of $\Delta T\sim 10\,000$--$12\,000\rm\,K$
(\citealt{McQuinn2009,Compostella2013}) and $\gamma-1 \sim
0.2$--$0.3$.  The volume of the IGM which is photoheated to
significantly higher temperatures than this is generally expected to
be small.  Furthermore, many of the hard photons may deposit their
energy in dense regions which will cool rapidly and are not probed
by the \lya forest measurements (\citealt{Bolton2009a}). These high
column density systems are currently not well captured in radiative
transfer simulations of He\,\textsc{ii} reionization, and must be
accounted with sub-grid models or a global clumping factor for the
gas.

On the other hand, where homogeneous UVB models break down is correctly
modelling the patchy nature of the heating during He\,\textsc{ii}
reionization. The inhomogeneous heating of the IGM will lead to
significant spatial variations in the flux and spectral shape of the
UVB, and most likely also substantial temperature
fluctuations. Note, however, observational evidence for the latter
appears to be difficult to obtain from line-of-sight \lya forest
measurements \citep{Theuns2002,Lai2006,McQuinn2011}. One should keep in mind though that some
of the measurements of the IGM temperature, in particular those that are based on the lower cutoff of the
line width distribution, are susceptible to being biased low in the presence of spatial temperature
fluctuations. Despite the good agreement between IGM temperatures in optically-thin simulations and observations, we thus cannot rule out the possibility that both are biased low to some extent. Spatially fluctuating heating will certainly lead to an increased scatter in the
temperature-density plane that is not captured well in simulations with a homogeneous UVB. It might also bias the mean temperature at some level. Such models are therefore still not a substitute for performing full radiative transfer calculations in
large volumes which capture these effects. As already discussed, there may furthermore be
additional heating processes (such as e.g. the TeV blazar heating we
have already briefly mentioned) which are not accounted for in the
simulations.

However, the surprising success of our modelling presented here is
attributable to two reasons.  Firstly, it requires capturing the
timing of the reionization of He\,\textsc{ii} as measured by the
He\,\textsc{iii} volume filling factor, which in the HM2012 model is
controlled by the assumed ionizing emissivity due to quasars and the
modelling of the spatially averaged number of recombinations based
on the clumping factor description of \citet{Pawlik2009}.  Secondly,
the spatial averaging performed in HM2012 to calculate
photoionization and photoheating rates accounts -- at least in a
volume averaged sense -- for the transition from optically thick to
thin heating.  This results in a thermal history in good agreement
with measured temperatures from the \lya forest. 

\section{Summary and conclusions}
\label{sec:conclusions}

We have performed here cosmological hydrodynamical simulations with a
non-ionization-equilibrium version of \mbox{\textsc{p-gadget3}} and
the HM2012 UVB flux. We carefully compare the thermal state
of the IGM, as well as H\,\textsc{i} and He\,\textsc{ii}
Lyman-$\alpha$ forest opacities, to the latest observational
constraints. Our main results are as follows:

\begin{itemize}[leftmargin=5mm]

\item The IGM temperature in the simulations are in good agreement
  with recent observational constraints.  The agreement becomes
  excellent once we correct for the timing of He\,\textsc{ii}
  reionization based on the volume filling factor predicted by
  spatially averaged emissivities and recombination rates assumed in
  HM2012. The predicted IGM temperature at $z \lesssim 3$ is somewhat
  lower than observed. This may suggest that either photoheating is
  more efficient than in our simulations or alternatively leaves room
  for not yet accounted additional heating processes, like, e.g.,
  heating by TeV blazars (\citealt{Puchwein2012}). 

\item Our numerical simulations predict He\,\textsc{ii} Lyman-$\alpha$ forest
  opacities that are somewhat lower than observed for $2.5 \lesssim z
  \lesssim 3.3$. Taking the spatial variations
  expected at the tail-end of He\,\textsc{ii} reionization into
  account will also be important.  On the other hand, at $z \gtrsim
  3.3$, our predicted He\,\textsc{ii} opacities are significantly
  larger than the measurements by \citet{Worseck2014} (see also
  \citealt{Compostella2014}).  If these new data are confirmed with
  further observations, this may suggest there is significant tension
  between the measured evolution of temperature and He\,\textsc{ii}
  opacity that merits further investigation.

\item The effective optical depth of the hydrogen
  Lyman-$\alpha$ forest predicted by our simulations at redshifts $2.5 \lesssim z \lesssim 4$
  matches observations well. However, we confirm that at lower and
  higher redshifts, the optical depth is overpredicted. This suggests
  that the photoionization rate in the HM2012 model is too low at both
  $z \lesssim 2.5$ and $z \gtrsim 4$.

\item A comparison of our equilibrium and non-equilibrium simulations
  corroborates previous findings that non-equilibrium effects are
  indeed significant, even when modelling photoheating with a homogeneous UVB. They, thus, ideally need to be taken
  into account in cosmological hydrodynamical simulations as standard.

\end{itemize}

\noindent
Finally, we remark that the good overall agreement of our simulations
with the data is encouraging, as it suggests that with some further
modest adjustments to the emissivities and mean free paths in the
HM2012 model, it should be possible to obtain a physical model which
allows faithful forward modelling of the \lya forest with
hydrodynamical simulations that is in agreement with both observed
temperatures and \lya opacities. This should render the ad hoc
adjustments of the heating rates used in the past for many
applications unnecessary.

\section*{Acknowledgements}

We would like to thank Joseph Hennawi, Gabor Worseck, Matthew McQuinn and Michael Shull
for helpful discussions, as well as Volker Springel for making \textsc{p-gadget3} available to us. We also thank Elisa Boera for providing us with the low 
redshift IGM temperature measurements. Support by the FP7 ERC Advanced Grant
Emergence-320596 is gratefully acknowledged. The simulations used in
this work were performed on the Darwin and Cosmos supercomputers at
the University of Cambridge. Part of the computing time was awarded
through the STFC's DiRAC initiative. JSB acknowledges the support of
a Royal Society University Research Fellowship. Support for this work was also 
provided by the NSF through grants OIA-1124453 and AST-1229745 and by NASA through grant NNX12A587G (PM).

\bibliographystyle{mn2efixed}
\bibliography{ref}

\appendix

\section{Non-equilibrium effects on the thermal and ionization history of the IGM}
\label{sec:non-eq_effects}

Non-ionization-equilibrium effects play an important role for the
photoheating and thus the thermal state of the low-density IGM
\citep[e.g.][]{Theuns1998,Hui1997}. In this appendix, we will discuss
these effects in detail.

\subsection{Integration of the rate equations}
\label{sec:integ_rate_eq}

The system of rate equations that we integrate in our non-equilibrium
code is of the form $\dot{\vec{y}} = f(\vec{y})$ where $\vec{y}$ is a
vector with the independent variables that we use, i.e. $\vec{y} = (u,
n_{\rm e}, n_{\rm HI}, n_{\rm HII}, n_{\rm HeI}, n_{\rm HeII}, n_{\rm
  HeIII})$.  Here $u$ is the specific internal energy and the other
variables are the individual abundances of free electrons, as well as
all of all ionization states of hydrogen and helium. The function $f$
is determined by our choice of rate coefficients (see
Sec.~\ref{sec:eq}). Note, that during the integration with the
\textsc{CVODE} library, we consider the individual abundances as
independent variables. The number conservation of electrons and
hydrogen and helium nuclei is then used as an independent check of the
integration accuracy during an individual gravity/hydrodynamic
timestep. At the end of each gravity/hydrodynamic timestep, we
renormalize the values to restore exact conservation. We adopt the
same error tolerance in the integration as \citet{Oppenheimer2013},
i.e. a relative error tolerance of $10^{-7}$ for the abundances of
the different ionization states of hydrogen and helium, as well as for
the free electron abundance. We also use \textsc{CVODE}'s Backward
Differentiation Formula scheme and Newton iteration.

As an additional test, we have compared \textsc{CVODE}'s solution for
a gas particle at mean cosmic density to the results of an explicit
integration of the rate equations with an extremely large number of
timesteps. The results are in excellent agreement.

\subsection{The IGM temperature}
\label{sec:temp_ne}

Figure~\ref{fig:T0} compares the temperature at mean density, or more
precisely $T_\textrm{median}(\Delta=1)$, between different
simulations, as well as to observations. It can be clearly seen that
the non-equilibrium treatment results in a much larger temperature
increase during the almost simultaneous H\,\textsc{i} and He\,{\sc i}
reionization between redshifts $\sim 15$ and $\sim 12$, as well as
during He\,\textsc{ii} reionization between redshifts $\sim 5$ and
$\sim 3.5$.

In the equilibrium computation, the heating rates are biased low as
they are directly proportional to the H\,\textsc{i}, He\,{\sc i} and
He\,{\sc ii} abundances. More precisely the photoheating rate per
volume is given by \citep[see e.g.][]{Katz1996}
\begin{equation}
  \mathcal{H} = n_{\rm HI} \mathcal{H}_{\rm HI} + n_{\rm HeI} \mathcal{H}_{\rm HeI} + n_{\rm HeII} \mathcal{H}_{\rm HeII},
\label{eq:heating_rate}
\end{equation}
where $n_{\rm HI}$, $n_{\rm HeI}$ and $n_{\rm HeII}$ are the particle
number densities for the different ionization
states. $\mathcal{H}_{\rm HI}$, $\mathcal{H}_{\rm HeI}$ and
$\mathcal{H}_{\rm HeII}$ are the photoheating rates per particle,
which depend only on the UVB. We use the values given
in Table~3 in HM2012. Thus, any underestimate of the neutral hydrogen
and helium or singly ionized helium abundance will results in an
underestimate of the photoheating rate and consequently of the IGM
temperature.

In the equilibrium treatment, an increase in the photoionization rates
results in an unrealistic instantaneous increase in the ionized
fractions as they are directly set to the new equilibrium values. In
the non-equilibrium calculation instead, it takes a while until enough
neutral or singly-ionized atoms are photoionized and the new
equilibrium state is approached. Thus, during reionization the degree
of ionization will be overestimated in the equilibrium
calculation. The corresponding underestimate of the neutral or
singly-ionized fractions is illustrated in
Figure~\ref{fig:HI_HeII_fracs}. It shows how the neutral hydrogen and
the He\,{\sc ii} fraction evolve as a function of redshift, both in
the equilibrium and the non-equilibrium calculation.

As expected the H\,{\sc i} fraction is underpredicted during hydrogen
reionization in the equilibrium model. This results in an
underestimate of the photoheating of hydrogen. The He\,{\sc i}
fraction is also biased low during He\,{\sc i} reionization, while the
He\,{\sc ii} fraction is overestimated under the assumption of
ionization equilibrium. As $\mathcal{H}_{\rm HeI} \gg \mathcal{H}_{\rm
  HeII}$ in the HM2012 model, the photoheating of helium is also
underpredicted in the equilibrium treatment. Together, this explains
the difference in IGM temperature during and after H\,{\sc i} and
He\,{\sc i} reionization between the non-equilibrium and equilibrium
simulation.

At $z \approx 11$, the ionization fractions are back in equilibrium
even in the non-equilibrium run. It takes, however, until $z \approx
7$ for the temperature difference to disappear.  At $z < 5$, a similar
effect can be seen due to He\,{\sc ii} reionization. The He\,{\sc ii}
abundance and the implied photoheating are underpredicted in the
equilibrium model. A much larger temperature boost is observed in the
non-equilibrium calculation. It takes until $z \approx 1$ for the
equilibrium and non-equilibrium IGM temperatures predictions to get
back into agreement.

\subsection{The slope of the temperature-density relation}
\label{sec:slope_ne}

In the following, we will discuss how the temperature-density relation
differs between the equilibrium and non-equilibrium simulations. The
logarithmic slope $\gamma - 1$ of the temperature-density relation at
mean density is shown in Fig.~\ref{fig:gamma_minus_1}. Details about
how the slopes are measured from the simulations are given in
Appendix~\ref{sec:measure_eos}.

\begin{figure}
\centerline{\includegraphics[width=\linewidth]{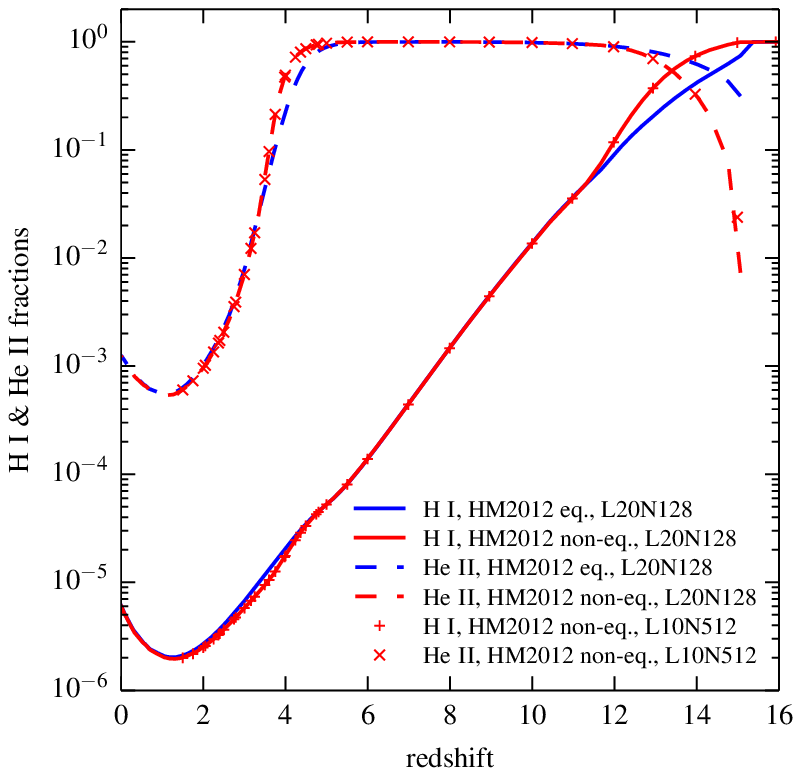}}
\caption{Ionization fractions as a function of redshift for
  simulations with equilibrium and non-equilibrium photoheating. The
  results are based on the HM2012 UVB. Shown are the ratios
  of the number of H\,\textsc{i} to the number of all hydrogen atoms
  and of the number of He\,\textsc{ii} to all helium atoms. The
  L20N512 simulation is in excellent agreement with the other runs and
  not shown for clarity.}
\label{fig:HI_HeII_fracs}
\end{figure}

During H\,\textsc{i} and He\,\textsc{i} reionization the
temperature-density relation is almost isothermal in the
non-equilibrium simulation. The reason for this is that regions of
different density are photoheated by the same spectrum in our
simulations with a homogeneous UVB. This results in a roughly constant
temperature during and shortly after reionization. In the equilibrium
run instead, the amount of heating is proportional to the equilibrium
neutral fraction, which is higher in high density regions due to the
larger recombination rate. As a consequence, the temperature-density
relation quickly attains a positive slope. In the non-equilibrium
calculation, the difference in recombination rate only becomes
important once it becomes comparable to the photoionization rate,
i.e. once ionization equilibrium is approached. This happens around
redshift 12 (see Fig.~\ref{fig:HI_HeII_fracs}). From that point on the
temperature-density relation also steepens in the non-equilibrium run,
mostly by a decrease of the temperature in low-density regions in
which the photoheating can no longer offset the inverse-Compton
cooling by the cosmic microwave background. Note, that in both cases,
the slope of the temperature-density relation during H\,\textsc{i} and
He\,\textsc{i} reionization is not mainly set by adiabatic compression
and expansion, but by the difference in the effectiveness of
photoheating.

As expected and as discussed in Sec.~\ref{sec:T-rho}, the temperature
boost during He\,\textsc{ii} reionization translates into a
significant flattening of the temperature-density relation in the
non-equilibrium simulation. The flattening is stronger in regions that
have a lower initial temperature, i.e. in regions with a lower
density.

In the equilibrium simulation, instead, we do not observe a large
change in the logarithmic slope, nor a steepening of the
temperature-density relation with increasing density. This can be
understood in the following way. Reionization proceeds very quickly in
the equilibrium computation as an increase in the photoionization rate
results in an unrealistic instantaneous increase in the ionized fraction. The photoheating
is, however, not directly coupled to the change in the ionized
fraction but happens with some delay, i.e. at a time when the IGM is
already largely ionized. According to Eq.~(\ref{eq:heating_rate}) the
heating rate is given by $\mathcal{H} \approx n_{\rm HeII}
\mathcal{H}_{\rm HeII}$, where we have ignored the subdominant
contribution from H\,\textsc{i} and He\,\textsc{i} during
He\,\textsc{ii} reionization. Next, we note that the He\,\textsc{iii}
recombination rate in the relevant range is roughly proportional to
$\propto T^{-0.7}$. Thus, when also ignoring collisional ionization,
which is not important at low density, ionization equilibrium
corresponds to a balance of photoionization and recombination,
i.e. $n_{\rm HeII} \Gamma_{\rm HeII} \propto n_{\rm HeIII} n_{\rm e}
T^{-0.7}$. Once He\,\textsc{ii} is mostly ionized, we have $n_{\rm
  HeIII} \propto n_{\rm e} \propto \rho$. Therefore, the heating rate satisfies the
following proportionality relation, $\mathcal{H} \approx n_{\rm HeII}
\mathcal{H}_{\rm HeII} \propto \rho^2 T^{-0.7}$. After a period of
heating the temperature change is then proportional to $\Delta T
\propto \mathcal{H} / \rho \propto \rho T^{-0.7} \propto
\rho^{1-0.7(\gamma-1)} \propto \rho^{1.7-0.7 \gamma}$, where we have
assumed that the initial temperature-density relations has a
logarithmic slope $\gamma-1$. The relative change hence satisfies
$\Delta T / T \propto \rho^{1.7-0.7 \gamma - \gamma + 1} \propto
\rho^{2.7-1.7 \gamma}$, so that for an initial value of $\gamma -1
\approx 2.7/1.7 - 1 \approx 0.59$ the slope of the temperature-density
relation does not change by photoheating when followed under the
assumption of ionization equilibrium. As the slope before
He\,\textsc{ii} reionization is quite close to this value, no
significant change of the slope is observed.

Note that as the recombination rate of hydrogen is also roughly
$\propto T^{-0.7}$, a similar calculation holds for the photoheating
after hydrogen reionization. The photoheating, thus, pushes the
temperature-density relation towards the \textit{stable} slope of
$\approx 0.59$, thereby explaining the much quicker increase of
$\gamma - 1$ at redshifts 15 to 12 compared to the non-equilibrium
simulation (see Fig.~\ref{fig:gamma_minus_1}).

As a final remark, we would like to point out that the relatively flat
slope of the temperature-density relation at $\Delta=10^{0.5}$ and $z
\gtrsim 6$ (as shown by the \textit{dotted} curves in
Fig.~\ref{fig:gamma_minus_1}) is a consequence of radiative
cooling. Such a flattening at high density can also be seen in
Fig.~\ref{fig:rhoT} for $z=3.5$ and $\Delta \gtrsim 10$. Due to the
larger value of the mean density at higher redshift this becomes
important at lower $\Delta$ values there.

\section{The local absorption approximation}
\label{sec:loc_absorp}

For the immediate local absorption approximation (discussed in Sec.~\ref{sec:hmuvb}), the excess energy is
computed assuming that all emitted ionizing radiation above the
ionization threshold with a mean free path shorter than the Hubble
radius is absorbed. The latter criterion translates to a high-energy
cut-off to the UVB spectrum which decreases toward
lower redshift. This cut-off is relevant only for the He\,\textsc{ii} excess energy, which is, however, also fairly
insensitive to its exact value as long as the $\sim 30$ keV bump in the HM2012 quasar emissivity is excluded. 
For simplicity, we, thus, derive the mean free path that
enters the computation of the cut-off energy for a homogeneous
universe with the same ionization fractions as our non-equilibrium
runs. 

If there were only one single species of absorbers, it would
be sufficient to weight the emitted spectrum with a constant
weight between the ionization threshold and the high-energy cutoff and
compute the mean excess energy. However, for multiple species -- we
consider absorption by H\,\textsc{i}, He\,\textsc{i} and
He\,\textsc{ii} -- one has to keep track of what fraction of the
radiation is absorbed by each species in order to obtain a mean excess
energy for each species separately. We do this by assuming that the
fraction absorbed by each species at a specific wavelength is
proportional to the product of the species' number density and its
photoionization cross section at that wavelength. For the number
densities we use the ionization fractions in the non-equilibrium
simulations (see Fig.~\ref{fig:HI_HeII_fracs}).

Note that in Fig.~\ref{fig:uvb} the local absorption estimates are truncated at the redshift where
the upper energy cut-off adopted for the spectrum becomes less than twice the ionization threshold.

\section{HEII absorbers in the Haardt \& Madau 2012 model}
\label{sec:heii_absorbers}

The thermal evolution of the IGM during He\,\textsc{ii} reionization is obviously sensitive to the spectrum of the UV background by which it is ionized. In this work we employ the HM2012 UVB model. In the following, we discuss some of the uncertainties in this model and how they affect the thermal evolution during He\,\textsc{ii} reionization. Critical ingredients in predicting the He\,\textsc{ii} ionizing background are the spectra of the ionizing sources, as well as the spectral filtering by the intervening IGM. In the HM2012 model, the latter is described by an empirical absorber H\,\textsc{i} column density distribution and a prescription for converting the H\,\textsc{i} column density $N_{\rm HI}$ of an absorber to its He\,\textsc{ii} column density $N_{\rm HeII}$. The opacity of the He\,\textsc{ii} is then taken into account when integrating the evolution equation of the UVB. The H\,\textsc{i} column density distribution is constrained rather well from observations at the redshifts relevant for He\,\textsc{ii} reionization \citep[see e.g.][]{O'Meara2013,Kim2013,Rudie2013}. In the remainder of this section, we thus focus on the uncertainties in the $N_{\rm HI}$ to $N_{\rm HeII}$ conversion.

In HM2012, this conversion is based on a radiative transfer calculation in which absorbers are treated as semi-infinite slabs which are illuminated by the external UVB. The $N_{\rm HeII}/N_{\rm HI}$ ratio obtained in this way of course depends on the spectrum of the UVB, as well as on the assumed thickness of the absorber. In the HM2012 calculation the UVB and the absorber properties are coupled self-consistently, as the opacity of the absorbers is taken into account when evolving the UVB. The main additional assumption that is required concerns the thickness of the absorber. HM2012 assume that the absorber size is given by the Jeans scale, which can be theoretically motivated for overdense absorbers in local hydrostatic equilibrium \citep{Schaye2001}. We now explore how sensitive results are to this assumption.

\begin{figure*}
\centerline{\includegraphics[width=\linewidth]{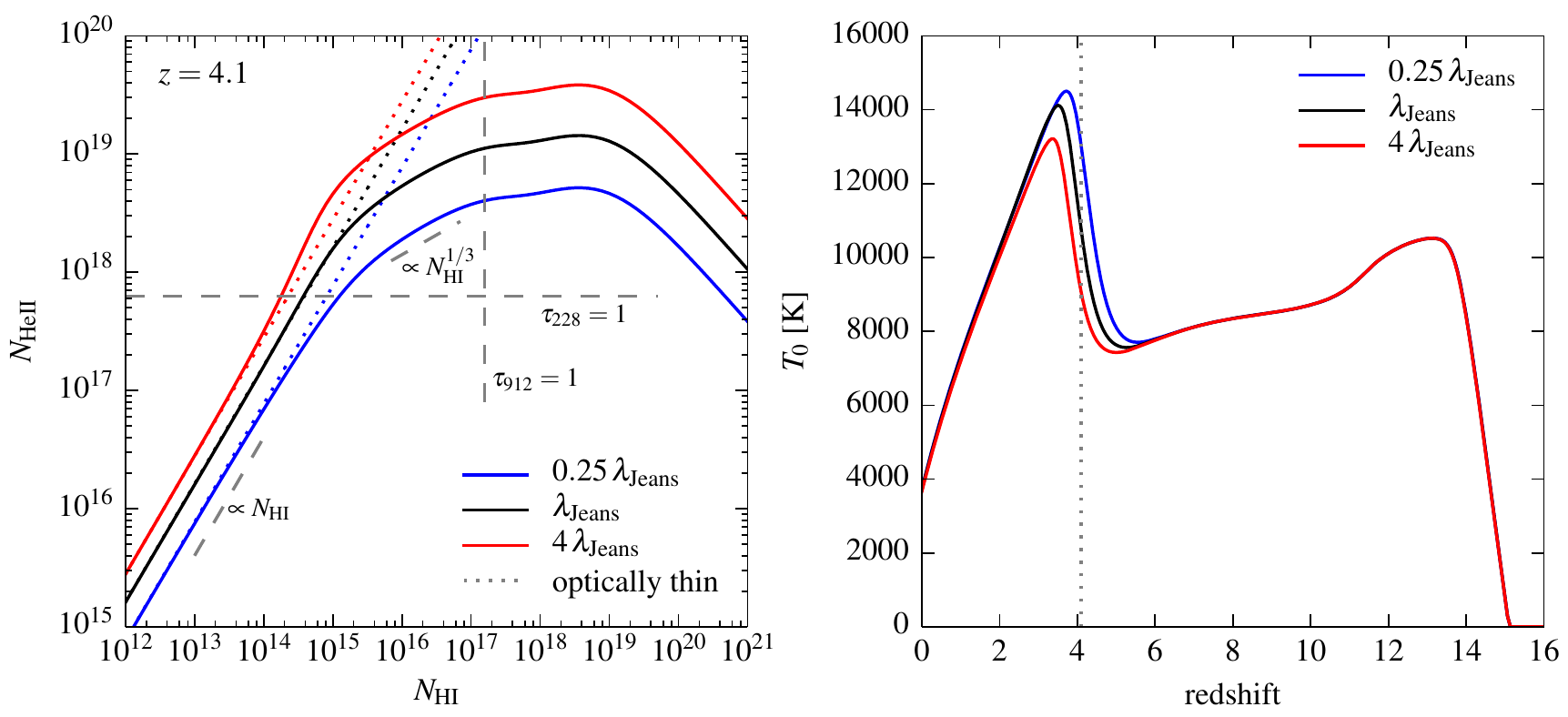}}
\caption{The \textit{left} panel illustrates the conversion from the H\,\textsc{i} to the He\,\textsc{ii} column density of an absorber in the HM2012 model. In particular it shows how the conversion depends on the assumed size of the absorber in units of the Jeans length. Results are shown for absorber sizes of 0.25, 1 and 4 times the Jeans length at $z=4.1$. The \textit{right} panel shows how the evolution of the IGM temperature at mean density depends on the assumed absorber size. The curves are based on a non-equilibrium treatment of photoheating.}
\label{fig:absorber_size}
\end{figure*}

The left panel of Fig.~\ref{fig:absorber_size} shows the $N_{\rm HeII}$-$N_{\rm HI}$ relation in HM2012 (black curve) during He\,\textsc{ii} reionization at $z=4.1$. Also shown are results for 4 times thicker (red) and 4 times thinner (blue) absorbers. These changes in absorber size result in factor of 2 to 3 changes in He\,\textsc{ii} column density. The increased absorption by He\,\textsc{ii} when assuming larger absorbers results in a softer UVB which slightly delays He\,\textsc{ii} reionization. Due to the then lower excess energy per ionization event this also somewhat decreases the temperature boost as illustrated in the right panel of Fig.~\ref{fig:absorber_size}, which shows the evolution of the IGM temperature at mean density for the three different assumed values of absorber size. Despite the significant changes in absorber size, the effect on the thermal history of the IGM is rather small compared to the difference between an equilibrium and a non-equilibrium treatment of photoheating. 

In reality the patchy nature of He\,\textsc{ii} reionization will result in scatter in the properties of the ionizing background to which absorbers are exposed. This translates into scatter in the $N_{\rm HeII}$-$N_{\rm HI}$ relation which might further modify the thermal evolution. In a homogeneous UVB model like HM2012, this cannot be followed faithfully. Fig.~\ref{fig:absorber_size} at least gives some idea how sensitive the thermal evolution is to changes in the $N_{\rm HeII}$ distribution.  

For those readers interested in the details of the $N_{\rm HI}$ to $N_{\rm HeII}$ conversion in HM2012, we will in the remainder of Appendix~\ref{sec:heii_absorbers} shed some light on the processes that shape the relation between the column densities. In the optically thin limit, at low column densities, $N_{\rm HI}$ and $N_{\rm HeII}$ are simply proportional to each other (see the discussion in HM2012). The proportionality constant depends on the hardness of the UVB. The different optically thin values of $N_{\rm HeII}$ in the left panel of Fig.~\ref{fig:absorber_size} (also shown by the dotted curves) are a consequence of the different UVB implied by the self-consistent coupling of absorbers and UVB evolution. For larger absorbers the softer UVB translates into larger $N_{\rm HeII}$ values in this regime.

At higher column density when He\,\textsc{ii} becomes optically thick (indicated by the horizontal dashed line), the $N_{\rm HeII}/N_{\rm He}$ ratio approaches unity, where $N_{\rm He}$ is the total helium column density independent of ionization state. Depending on absorber size the $N_{\rm HeII}/N_{\rm He}$ ratio starts out from different optically thin values. This translates to different slopes of the $N_{\rm HeII}$-$N_{\rm HI}$ relations during the transition to $N_{\rm HeII}/N_{\rm He} \approx 1$. In Fig.~\ref{fig:absorber_size}, this happens in the range $10^{14} \lesssim N_{\rm HI} \lesssim 2 \times 10^{15}$. For even higher column densities ($N_{\rm HI} \gtrsim 2 \times 10^{15}$) $N_{\rm HeII}$ is roughly proportional to $\propto N_{\rm HI}^{1/3}$. This corresponds to $N_{\rm HeII}/N_{\rm He} \approx 1$ and $N_{\rm HI}/N_{\rm H}$ still being in the optically thin regime, so that $N_{\rm HeII} \propto N_{\rm He} \propto N_{\rm H}$ and $N_{\rm HI} \propto N_{\rm H} n_{\rm e} \propto N_{\rm H} N_{\rm HI}^{2/3}$, where the electron density $n_{\rm e}$ affects the number of recombinations to H\,\textsc{i} and the proportionality $n_{\rm e} \propto N_{\rm HI}^{2/3}$ assumes a constant size of the absorber in units of the Jeans length (see equation 30 in HM2012). It then follows that $N_{\rm HeII} \propto N_{\rm H} \propto N_{\rm HI}^{1/3}$. This holds until H\,\textsc{i} also becomes optically thick at $N_{\rm HI} \approx 10^{17} {\rm cm}^{-2}$. At this point, a large increase in $N_{\rm HI}$ is caused by a small increase in $N_H \propto N_{\rm HeII}$ so that the $N_{\rm HeII}$-$N_{\rm HI}$ becomes very flat.

At even larger column density, He\,\textsc{i} becomes optically thick as well. In Fig.~\ref{fig:absorber_size} this happens between $N_{\rm HI} \approx 10^{18} {\rm cm}^{-2}$ and $10^{19} {\rm cm}^{-2}$ and results in helium becoming increasingly neutral. This effect alone would result in $N_{\rm HeII}$ levelling off. However, hydrogen also becomes largely neutral at these column densities so that $N_{\rm HI} \approx N_{\rm H}$. For an absorber with a size that is a fixed multiple of the Jeans scale, a further increase in $N_{\rm HI}$ then correspond to an increase in particle number density which boosts recombination rates. As a consequence $N_{\rm HeII}$ decreases with a further increase in $N_{\rm HI}$.

\section{The connection between spectral curvature and the flux power spectrum}
\label{sec:curv_vs_powspec}

\citet{Becker2011} use the mean curvature of the Lyman-$\alpha$
absorption spectra as a proxy for the IGM temperature. They define the
curvature by
\begin{equation}
  \kappa = \frac{ \frac{\textrm{d}^2F}{\textrm{d}v^2} }{\left[1 \, \textrm{km}^{-2} \textrm{s}^{2} + (\frac{\textrm{d}F}{\textrm{d}v})^2\right]^{3/2} },
\label{eq:def_kappa}
\end{equation}
where $F$ is the transmitted flux fraction and $v$ is the velocity
offset. To obtain temperatures, simulations are used to calibrate a
relation between the mean of the absolute value of $\kappa$,
i.e. $\langle | \kappa| \rangle$, and the temperature at a
characteristic overdensity (see also Sec.~\ref{sec:temp_defs}). The
mean is calculated for all pixels with $0.1 < F < 0.9$. This relation
can then be used to translate the curvature of an observed spectrum to
an IGM temperature.

We would like to better understand what spatial scales dominate the
mean curvature. To this end, we try to relate it to the flux power
spectrum. This is possible when using three simplifications:
\begin{itemize}[leftmargin = 7mm]

\setlength{\itemindent}{0.3cm}

\item Employing a root mean square average of $\kappa$ rather than the
  mean of the absolute value.

\item Including all pixels, i.e. also those with $F<0.1$ and $F>0.9$.

\item Assuming $(\frac{\textrm{d}F}{\textrm{d}v})^2 \ll 1 \,
  \textrm{km}^{-2} \textrm{s}^{2}$ in the denominator of
  Eq.~(\ref{eq:def_kappa}). This is typically well
  satisfied. Neglecting the $(\frac{\textrm{d}F}{\textrm{d}v})^2$ term
  does, thus, not change the value of $\kappa$ significantly.

\end{itemize}
The root mean square value of $\kappa$ is then given by
\begin{equation}
  \langle \kappa \rangle_{\rm RMS} = \sqrt{\frac{1}{N} \sum_{n=0}^{N-1} \kappa_{n}^2} = \sqrt{\frac{1}{N^2} \sum_{l=0}^{N-1} \hat{\kappa}_{l}^2},
\label{eq:kappa_rms}
\end{equation}
where $n$ is the pixel index and $N$ is the number of pixels in the
spectrum. In the second equality we use Parseval's theorem to rewrite
the mean curvature in terms of the discrete Fourier transform (denoted
by $\hat{}$ ) of $\kappa$. Neglecting the
$(\frac{\textrm{d}F}{\textrm{d}v})^2 \ll 1 \, \textrm{km}^{-2}
\textrm{s}^{2}$ in the denominator of Eq.~(\ref{eq:def_kappa}), this
can be easily related to the Fourier transform of $F$. We note that
\begin{equation}
  \hat{\kappa}_l \approx \hat{(\frac{\textrm{d}^2F}{\textrm{d}v^2})}_{l} = -k_{l}^2 \hat{F}_{l},
\end{equation}
where $k_l = 2 \pi / \Delta v \times \min(l,N-l)$ with $\Delta v$ being the length of the spectrum in velocity space. Using this we can rewrite Eq.~(\ref{eq:kappa_rms}) as
\begin{equation}
  \langle \kappa \rangle_{\rm RMS} \approx \sqrt{\frac{1}{N^2} \sum_{l=0}^{N-1} k_{l}^4 \hat{F}_{l}^2} =
  \sqrt{\frac{1}{\Delta v} \sum_{l=0}^{N-1} k_{l}^4 P_{l}},
\end{equation}
i.e. in terms of the flux power spectrum $P_{l} \equiv \Delta v \hat{F}_{l}^2 / N^2$.

\begin{figure}
\centerline{\includegraphics[width=\linewidth]{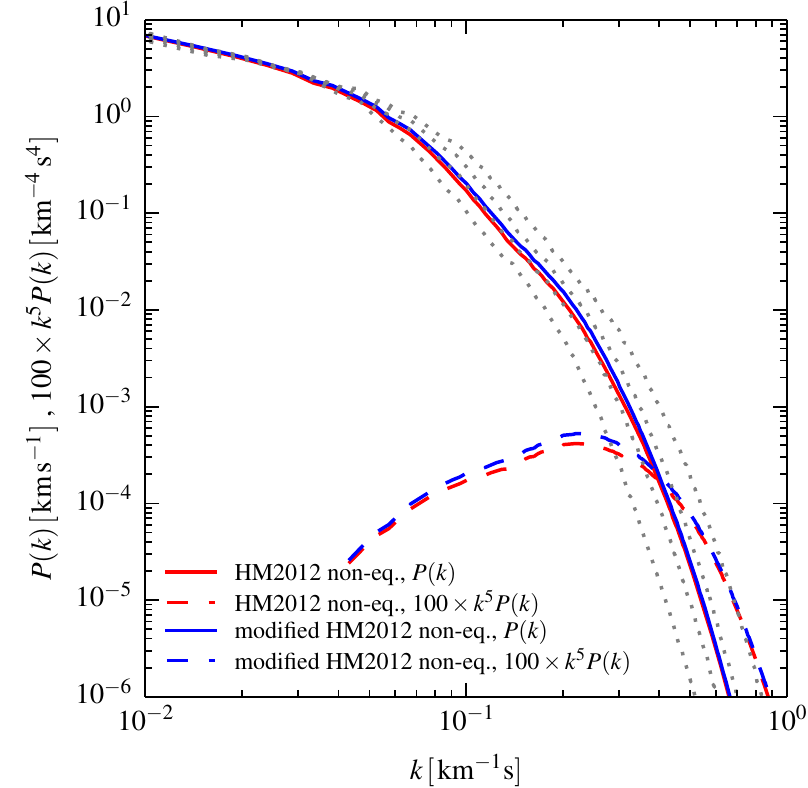}}
\caption{Flux power spectrum and dominating scales in the spectral
  curvature at $z=4.6$. The power spectra are shown for
  non-equilibrium simulations (L10N512) with the HM2012 background and
  the modified HM2012 background that is discussed in
  Sect.~\ref{sec:curv_temp}, as well as for some of the simulations
  that were used in \citet{Becker2011} to calibrate the relation
  between spectral curvature and IGM temperature (A15, AB15, B15, C15;
  \textit{top} to \textit{bottom, grey dotted} curves). The
  differential contribution to the spectral curvature
  $\textrm{d}\langle \kappa^2 \rangle / \textrm{d} \ln(k) \propto 100
  \times k^5 P(k)$ is indicated for the non-equilibrium simulations
  with the HM2012 and modified HM2012 backgrounds.  }
\label{fig:powspec}
\end{figure}

In other words the contribution of a specific scale to $\langle
\kappa^2 \rangle$ is proportional to $k_{l}^4 P_{l}$ or when writing
this in a continuous form $\propto k^4 P(k) \, \textrm{d}k = k^5 P(k)
\, \textrm{d}(\ln k)$. The latter quantity rescaled by a factor of 100
for clarity, i.e. $100 \times k^5 P(k) \propto \textrm{d}\langle
\kappa^2 \rangle / \textrm{d} \ln(k)$, is shown in
Fig.~\ref{fig:powspec} for non-equilibrium simulations with the HM2012
background and the modified HM2012 background. The latter was modified
such that H\,\textsc{i} and He\,\textsc{i} reionization happen
significantly later, while leaving the thermal state at $z<6$
unchanged (see Sect.~\ref{sec:curv_temp}). These two models have,
thus, the same instantaneous temperature, but the modified HM2012
background results in less Jeans smoothing, in particular for $z
\gtrsim 3.5$. Also shown are flux power spectra both for the two
models just described, as well as for some of the reference
simulations used in \citet{Becker2011}.

The figure illustrates that Jeans smoothing mostly affects
intermediate scales $0.05 \, \textrm{km}^{-1} \textrm{s} \lesssim k
\lesssim 0.4 \, \textrm{km}^{-1} \textrm{s}$. Larger scales are not
very sensitive to Jeans smoothing, but also not to instantaneous
temperature. The latter can be seen from the \textit{grey dotted}
curves, which all correspond to different normalizations of the
temperature-density relation. On small scales the flux power spectrum
has a weak dependence on the Jeans smoothing but a strong dependence
on the instantaneous temperature.

The largest contribution to the spectral curvature comes from the same
scales on which we find the largest sensitivity to the amount of Jeans
smoothing. This can be most easily seen by comparing the two
\textit{dashed} curves which have the same relative difference as the
power spectra, but show at the same time the scales that contribute
most to the curvature. This makes clear that spectral curvature
measures a combination of Jeans smoothing and instantaneous
temperature. It also suggests that using the full information in the
flux power spectrum, i.e. including smaller scales, may help to break
this degeneracy.

\section{Measuring the slope and normalization of the $\rho-T$ relation in simulations}
\label{sec:measure_eos}

We measure the slope and normalization of the temperature-density
relation at density $\Delta$ (in units of the mean baryon density) by
finding the mode of the volume-weighted $\log_\textrm{10}(T)$
distribution at fixed density at sampling points $\Delta_1 = \Delta /
1.25$ and $\Delta_2 = \Delta \times 1.25$. At both densities we have
to use a finite bin size to compute the mode from a suitably large
number of gas particles. More precisely, we use all gas particles with
densities within 5 percent of $\Delta_1$ or $\Delta_2$.

This can, however, slightly bias the value of the mode as one can
essentially end up with any temperature value on the ridge of the
temperature-density relation within the density range given by the
bin size. To avoid this problem, we scale the temperatures of all
particles within the bin to the bin centre using an initial guess of
the slope of the temperature-density relation. We then compute the
mode of the distribution of the rescaled logarithmic temperatures with
the \textit{half-sample mode estimator} \citep{Bickel2005}. This
yields the logarithmic temperatures $\log_\textrm{10} T_1$ and
$\log_\textrm{10} T_2$. The logarithmic slope $\gamma - 1$ is then
computed in a straightforward way by
\begin{equation}
 \gamma - 1 = \frac{\log_\textrm{10} T_2 - \log_\textrm{10} T_1}{\log_\textrm{10} \Delta_2 - \log_\textrm{10} \Delta_1}.
\end{equation}
 
\noindent
The disadvantage of this procedure is that the measured mode values
$\log_\textrm{10} T_1$ and $\log_\textrm{10} T_2$ mildly depend on the
initial guess for $\gamma - 1$. We, thus, repeat the procedure
described above iteratively until the value of $\gamma - 1$ has
converged. More precisely, we stop the iteration once $\gamma - 1$
changes by less than $10^{-6}$ in one iteration.

Using the final value of $\gamma - 1$, we scale the logarithmic
temperatures of all gas particles within 5 percent of density $\Delta$
to density $\Delta$. The normalization of the temperature-density
relation is then obtained by computing the mode of these rescaled
logarithmic temperatures. We refer to the temperature obtained in this
way also by $T_\textrm{mode}(\Delta)$.

By overplotting the measured temperature-density relations on
phase-space diagrams, i.e. plots similar to Fig.~\ref{fig:rhoT}, we
have confirmed that the method described here reliably recovers the
position and slope of the ridge of the temperature-density relation.

\section{Computing the thermal evolution from the HeII ionization history}
\label{sec:ion_history_to_T}

In Sec.~\ref{sec:min_ion_model}, we predict the thermal evolution during He\,\textsc{ii} reionization for a given evolution of the He\,\textsc{iii} volume filling factor $Q_{\textrm{He} \, \textsc{iii}}(z)$ based on Eq.~(\ref{eq:temp_heii_reion}) and the temperature in a simulation without He\,\textsc{ii} reionization $T_{0,\textrm{no HeIII}}$ (He\,\textsc{ii} photoheating and photoionization rates have been set to zero in this run). We additionally assume that the mean excess energy per ionization is well described by the HM2012 UVB model. For ion species $i$ it is, thus, given by
\begin{equation}
 E_i(z) = \frac{\mathcal{H}_i(z)}{\Gamma_i(z)},
\end{equation}
where $\mathcal{H}_i$ and $\Gamma_i$ are the photoheating and photoionization rates in the HM2012 model (see their Table~3). If He\,\textsc{ii} is newly reionized in a region, the temperature there increases by
\begin{align}
 \Delta T_{\textrm{He}\, \textsc{ii} \, \textrm{reion}}(z) = \, & T_{\textrm{0,no HeIII}}(z) \left( \frac{n_{\textrm{He} \, \textsc{ii} \, \textrm{region}}}{n_{\textrm{He} \, \textsc{iii} \, \textrm{region}}} - 1 \right) \nonumber \\
  & + \frac{E_{\textrm{He} \, \textsc{ii}} \, n_{\textrm{He}}}{\frac{3}{2} \, k \,\, n_{\textrm{He} \, \textsc{iii} \, \textrm{region}}},
\label{eq:T_heat}
\end{align}
where $k$ is the Boltzmann constant, $n_{\textrm{He}}$ is the number density of helium nuclei and $n_{\textrm{He} \, \textsc{ii} \, \textrm{region}}$ and $n_{\textrm{He} \, \textsc{iii} \, \textrm{region}}$ are the total particle number densities in He\,\textsc{ii} and He\,\textsc{iii} regions, respectively. For the assumed hydrogen mass fraction of $0.76$ the number density ratios are hence given by $n_{\textrm{He} \, \textsc{ii} \, \textrm{region}}/n_{\textrm{He} \, \textsc{iii}\, \textrm{region}} \approx 0.965$ and $n_{\textrm{He}}/n_{\textrm{He} \, \textsc{iii} \, \textrm{region}} \approx 0.035$. The first term on the right-hand side of Eq.~(\ref{eq:T_heat}) accounts for the temperature change at fixed thermal energy due to the increase in particle number, while the second term accounts for the energy deposition by photoionization.

In addition to the direct heating by photoionization by the advancing He\,\textsc{ii} reionization, we also need to account for the fact that cooling and heating rates in He\,\textsc{ii} and He\,\textsc{iii} regions differ even when the volume filling factor does not change. This difference arises due the temperature and particle number density dependence of these rates. Accounting for it has only a minor effect on the temperature increase during He\,\textsc{ii} reionization but is critical for getting the correct evolution, i.e. thermal asymptote, afterwards.

Note that cooling and heating in He\,\textsc{ii} regions is already followed in the simulation without He\,\textsc{ii} reionization. We therefore do not have to worry about it when computing $\Delta T_{0}(z)$. For He\,\textsc{iii} regions instead cooling and heating is not followed correctly in the simulation, as the rates are computed based on $T_{0,\textrm{no HeIII}}(z)$ and the number densities in He\,\textsc{ii} regions. The difference in the heating due to recombinations and subsequent photoionization depends on the average recombination time at mean density. For ion species $i$ and region $j$ (being either a He\,\textsc{ii} or He\,\textsc{iii} region) it is given by
\begin{equation}
  t_{\textrm{rec},j,i}(z) = \frac{1}{n_{\textrm{e},j}(z) \alpha_i(T_j(z))},
\end{equation}
where $n_{\textrm{e},j}(z)$ is the electron number density in region $j$ and $\alpha_i$ the recombination rate coefficient of ion species $i$ which depends on temperature (see Sec.~\ref{sec:eq} for our choice of rate coefficients). The recombination-induced photoheating rate in region $j$ is then given by
\begin{equation}
  \Delta T_{\textrm{heat-cool},j} = \sum_{i} \frac{(E_i - \frac32 k f_\textrm{cool} T_j)}{\frac32 k} \frac{n_{i,j}}{n_j} \frac{\Delta t}{t_{\textrm{rec},j,i}(z)},
\label{eq:T-heat-cool}
\end{equation}
where the sum goes over species H\,\textsc{ii} and He\,\textsc{iii} for He\,\textsc{iii} regions and over H\,\textsc{ii} and He\,\textsc{ii} for He\,\textsc{ii} regions. The term $-\frac32 k f_\textrm{cool} T_j$ accounts for thermal energy loss by recombination radiation. As slower particles are more likely to recombine $f_\textrm{cool}$ is smaller than one. We assume a value of $f_\textrm{cool} = 0.5$ which is a good approximation for the relevant temperature range (see e.g. Chapters 5 and 6 of \citealt{Spitzer2007} for a detailed discussion). $n_{i,j} \Delta t / t_{\textrm{rec},j,i}$ is the number density of ion-electron pairs that recombine during the time $\Delta t$ corresponding to the current step in redshift. The number densities $n_{i,j}$ of ion species $i$ in region $j$ and $n_j$ of all particles in region $j$ are computed based on the assumed hydrogen mass fraction and the ionization state of the region. The temperature $T_j$ which affects $\alpha_i(T_j)$ and enters Eq.~(\ref{eq:T-heat-cool}) is given by $T_{0,\textrm{no HeIII}}$ in He\,\textsc{ii} regions and by $T_{0,\textrm{no HeIII}} + \Delta T_0 / Q_{\textrm{He} \, \textsc{iii}}$ in He\,\textsc{iii} regions. In the last term dividing by $Q_{\textrm{He} \, \textsc{iii}}$ converts the average overall temperature increase to the average temperature increase in He\,\textsc{iii} regions. As discussed above the computation here is only relevant for $\Delta T_{0}(z)$ due to its effect on He\,\textsc{iii} regions. In  particular, the simulation correctly accounts for the He\,\textsc{ii} regions. Thus, the recombination-induced photoheating rates derived above are weighted with $Q_{\textrm{He} \, \textsc{iii}}$ in the last term of Eq.~\ref{eq:temp_heii_reion} to get the contribution to the overall average. Finally, we also account for the difference in inverse-Compton cooling rates in $T_{\textrm{heat-cool},j}$. This is however a very minor effect at the considered redshifts.

\end{document}